\documentclass[%
aps,
 reprint,
 amsmath,amssymb,
prper,
floatfix
]{revtex4-2}

\usepackage{graphicx}% Include figure files
\usepackage{dcolumn}% Align table columns on decimal point
\usepackage{bm}% bold math
\usepackage{tikz}
\usepackage{pgfplots}
\usepackage[export]{adjustbox}
\pgfplotsset{compat=1.17}
\usepackage{cleveref}
\usepackage{rotating}
\usepackage{enumitem}
\usepackage{siunitx}
\DeclareSIUnit\bar{bar}

\usepackage{multirow}

\usepackage{xcolor}
\begin{document}

%\preprint{APS/123-QED}

\title{Testing the Validity of Embedding-Based Similarity and Clustering for Handwritten Physics Solutions}% Force line breaks with \\

\author{Maike Tauschhuber}
 \email{mtauschhuber@neuroo.de}
 \affiliation{%
Department of Information Technology and Electrical Engineering, ETH Zurich, 8092 Zurich, Switzerland
}%

\author{Gerd Kortemeyer}
 \email{kgerd@ethz.ch}
 \affiliation{%
Rectorate and AI Center, ETH Zurich, 8092 Zurich, Switzerland
}%
\altaffiliation[also at ]{Michigan State University, East Lansing, MI 48823, USA}

\date{\today}% It is always \today, today,

\begin{abstract}
Text embeddings are increasingly used in physics education research to organize, compare, and cluster large collections of written text. Their appeal is clear: once student responses have been mapped into a vector space, similarity comparisons and clustering become computationally inexpensive. However, in assessment contexts, the relevant question is not merely whether clusters can be produced, but whether the geometry of the embedding space preserves grading-relevant distinctions. We tested this premise using 992 handwritten student-problem solutions from a high-stakes engineering thermodynamics exam, transcribed into five textual representations and embedded using nine embedding mechanisms. We compared embedding similarity and embedding-based hierarchical clusters against human-assigned scores. Across models, representations, and clustering choices, embedding similarity showed a consistent but modest relationship to score similarity, and the resulting clusters were score-enriched but not score-equivalent. Experiments with a synthetic data set suggest that this may be due to embeddings behaving like novices when categorizing physics-problem solutions, that is, their similarity geometry is strongly influenced by surface features rather than conceptual, semantic structure. These findings suggest that state-of-the-art embeddings can support exploratory organization and human-in-the-loop review of physics solutions, but they do not provide an unsupervised basis for grading without external validation against the assessment construct of interest.

\end{abstract}

\maketitle

\section{Introduction}

\subsection{Assessment in physics}
Problem solving is central to learning physics across formative settings (homework, in-class practice, low-stakes quizzes) and summative assessments (midterms, finals)~\cite{larkin1979understanding,gok2010general,docktor14,ince2018overview,dufresne2004,laverty12b,fakcharoenphol2014physics,wieman2005transforming}. Meaningful assessment must capture characteristic features of physics reasoning: mathematical modeling that blends conceptual insight with equations, computation, and theoretical principles~\cite{uhden2012modelling,tuminaro2007elements,teodorescu2013new}; transfer to novel scenarios~\cite{singh2008assessing}; multiple legitimate solution paths~\cite{larkin,hull2013,docktor2016assessing}; and coordination of diagrams, equations, graphs, and prose~\cite{meltzer2005relation,kohl05,kohl2007strongly,nguyen2009students}. Focusing on a final numeric or symbolic answer alone is therefore oftentimes inadequate to get the full picture; instructors must examine the solution path to understand and support students’ reasoning~\cite{hull2013,docktor2016assessing}. Yet providing such path-focused feedback at scale is difficult in large introductory courses, given the expertise, time, and logistical demands involved~\cite{wilcox2014coupled}. Historically, expert judgment has been essential to evaluate not just end results, but the logical, conceptual, and mathematical competencies revealed in student work~\cite{reif1976,reif1995,hsu2004,hattie2008,alsalmani23}. While some aspects of solutions can be evaluated by computers, diverse approaches and error patterns typically require human interpretation~\cite{kashyd01,kortemeyer08,risley2001,stelzer2001,dufresne02,fredericks2007,richards2011,perdian2013,docktor2016assessing,burkholder2020}.

Handwritten solution paths remain the gold standard in timed high-stakes exams (as typesetting mathematics would add irrelevant overhead), and for a small number of such assessments, the grading effort might be manageable for small-enrollment courses. Outside formal exams, frequent formative evaluation can substantially enhance learning through timely, targeted feedback, but it is often underused because evaluation is labor-intensive~\cite{laverty12b,offerdahl2019formative}. While in principle, typesetting mathematics in computer-readable format would be possible in asynchronous take-home assessments, questions of academic integrity in remote and hybrid contexts add further constraints~\cite{clark2020testing}. Because students’ priorities are shaped by what is assessed and how credit is assigned, scalable approaches that illuminate reasoning while remaining trustworthy and feasible are a pressing need~\cite{gomez2025student}.

\subsection{Automated grading and human oversight}
To address workload, physics education has a long tradition of automated systems that assist instructors~\cite{kashy93,kashy95,warnakulasooriya2005learning,kortemeyer08,gladding15,gutmann2018}. Most platforms, however, constrain responses to multiple choice, ranking, numeric entries, or closed-form algebraic answers. Although such formats can be psychometrically sound~\cite{scott2006,pawl2013,kortemeyer2016psy}, they have been criticized for neglecting communication and presentation~\cite{stewart2010,davis2016} and for failing to capture the breadth of students’ reasoning~\cite{wilcox2014coupled}; they can also incentivize guessing and other unproductive behaviors, especially with multiple attempts~\cite{palazzo2010,gonulatecs2017,kortemeyer2015empirical}.

Even before large-scale generative models, researchers broadened the response space with short answers and explanations supported by AI techniques~\cite{vanlehn2010andes,nakamura2016}. The Andes tutoring system, for example, permits substantial freedom in constructing solutions and interprets diverse traces via logical solution graphs, an algebra subsystem, and structured feedback rather than language modeling~\cite{vanlehn2010andes,shapiro2005algebra}. Classifier-based approaches to scoring short physics responses have also shown practically useful agreement with human raters in interactive tutoring contexts~\cite{nakamura2016}.

Recently, large language models (LLMs), particularly generative Pre-trained Transformers~\cite{chatgpt,gpt4,gpt4o}, have demonstrated strong capabilities on academic tasks~\cite{meyer2023chatgpt} and across educational uses~\cite{kasneci2023chatgpt}. In physics education, work spans classroom integration~\cite{yeadon2024impact,sperling2024artificial,polverini2024understanding}, research support~\cite{tschisgale2023integrating,kieser2023educational,wulff2024physics}, and problem solving~\cite{kung2022,achiam2023gpt,kortemeyer23ai,lopez2024challenging,wang2024examining}, including performance above typical post-instruction averages on several concept inventories~\cite{kortemeyer2025multilingual}. LLMs can assist with generating new problems~\cite{kuchemann23} and show promise for grading free-form student work~\cite{wilson22,wan24,kortemeyer24aigrading,kortemeyer2024grading,liu2024ai}. A human-in-the-loop stance remains prudent: confidence filters can route only trustworthy AI decisions forward and defer uncertain cases to human graders~\cite{kortemeyer2025assessing}. This strategy reduces load while maintaining reliability traditionally associated with deterministic, closed-form systems.

Deep learning methods aim to learn such structure implicitly, but limitations remain. Frakn{\'o}i et~al.~\cite{fraknoi2023embedding} report that embeddings of variable-free arithmetic expressions can be dominated by surface properties rather than semantic relations, suggesting that core mathematical principles may not be reliably learned. A broader analysis argues that multi-head attention, positional encodings, and feed-forward layers play distinct roles across representational levels, with lower layers emphasizing patterns and upper layers semantics. Xin He~\cite{he2024mathematical} emphasizes hierarchical structure by combining semantic and structural features, using hesitant fuzzy sets to measure formula similarity~\cite{xu2014hesitant} and a language model for representation learning over literature graphs. Beyond standalone formulas, hierarchical encoders and decoders have been applied to math word problems, enriching token representations via dependency structure and generating solutions with tree-based decoders~\cite{lin2021hms}.

\subsection{Text embeddings}
A direct way to obtain similarities among student solutions would be to ask a large language model to compare pairs of responses. This may be attractive in principle, since such a model could compare solution strategies more flexibly than a fixed vector representation. The challenge is scaling. In grading, each solution can be evaluated once, so the number of expensive model calls scales linearly with the number of submitted solutions, $O(n)$. Similarity analysis is inherently pairwise: for $n$ solutions to a problem, there are
$
\frac{n(n-1)}{2}
$
unordered pairs. A direct LLM-based similarity analysis would therefore require $O(n^2)$ expensive model calls. For response sets of a few hundred students, this already amounts to tens of thousands of pairwise comparisons per problem, and the cost grows rapidly with additional problems, transcription conditions, or model variants. Embedding-based methods offer a computationally attractive alternative: each solution is processed once to produce a vector representation, after which all pairwise similarities can be computed cheaply using local vector operations such as dot products or cosine similarity. This computational argument motivates the use of embeddings as a scalable approximation to pairwise semantic comparison. It does not, however, establish that the resulting similarity geometry is valid for grading-related interpretation; that validity is the empirical question tested in this study.

Text embeddings have recently become attractive tools for physics education research because they provide a scalable way to compare large collections of written responses, explanations, interviews, or abstracts~\cite{wulff2025applying}. Embedding models are trained on large collections of text to place pieces of text that are used in similar contexts near one another in a high-dimensional vector space; once a text has been mapped into this vector space, distances between vectors can be used for retrieval, visualization, and clustering. This makes embeddings appealing in settings where manual qualitative comparison is expensive and where researchers seek to identify recurring patterns in large corpora~\citep{odden_etal_2024_embeddings_per,caramaschi2025analyzing}, characterize curricula~\cite{buzzell2025modern}, or code student responses~\cite{gili2025combining,wyrwich2025tracking}.

However, assessment contexts impose a stronger requirement than exploratory organization. A cluster of responses may be coherent in embedding space without being coherent with respect to the construct being assessed. In grading, the relevant construct is not general semantic similarity, but rubric-relevant correctness: the selection of appropriate principles, the coordination of representations, the validity of assumptions, the structure of derivations, and the treatment of numerical and symbolic details. Two solutions may be textually similar while differing in a crucial sign, state variable, unit, or physical assumption; conversely, two solutions may look quite different while earning similar scores because they instantiate different valid solution paths.

We therefore treat embedding-based clustering as a hypothesis to be tested rather than as a grading method. Following validity arguments in educational measurement, the question is not whether a computational representation can produce clusters, but whether the intended interpretation and use of those clusters are supported by evidence~\cite{kane_2013_validating,messick_1995_validity}. In the present study, we evaluate whether embedding similarity and embedding-based hierarchical clusters preserve human-assigned score similarity for handwritten physics solutions.

\begin{figure}
\begin{center}
\includegraphics[width=\columnwidth]{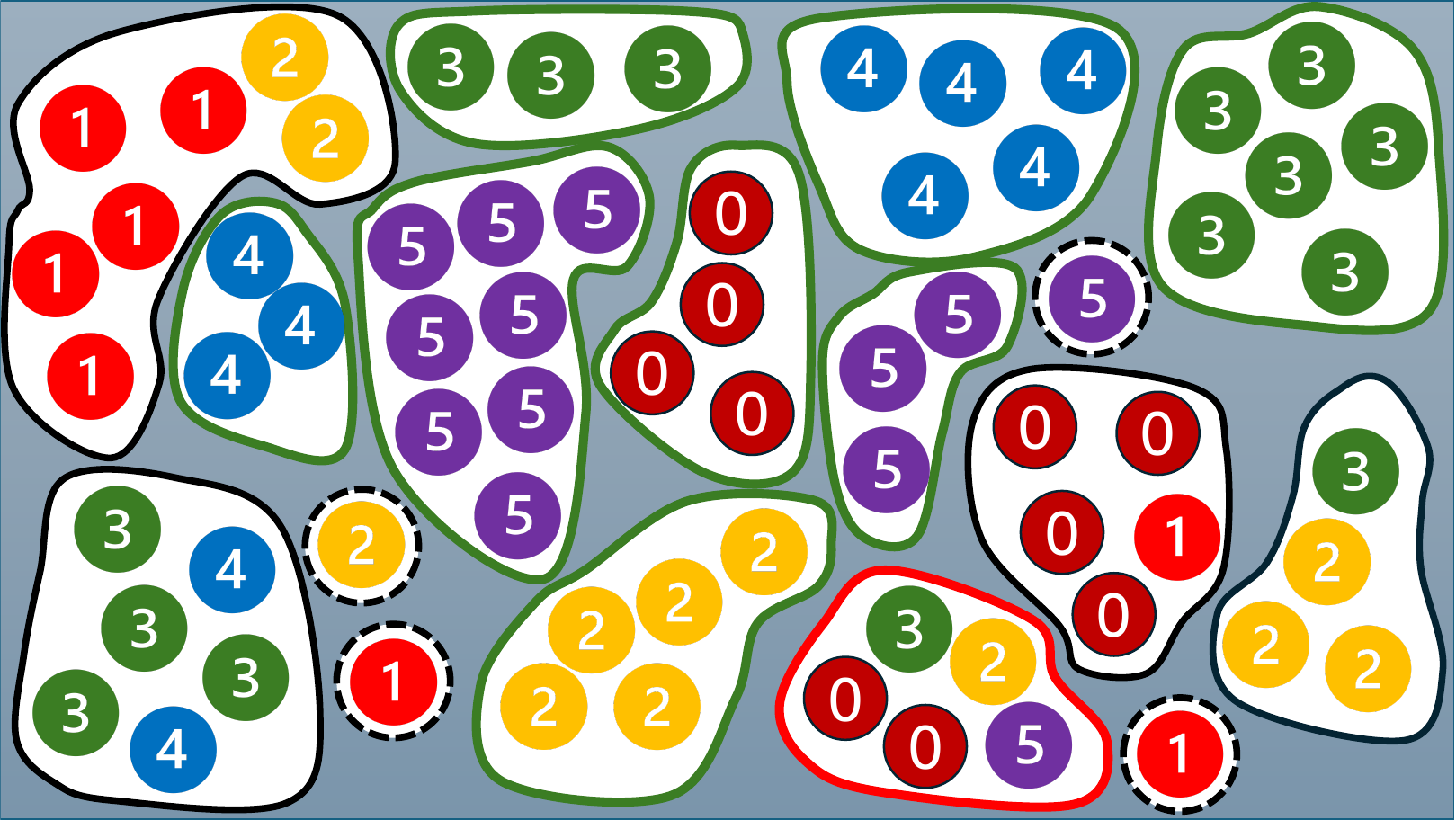}
\end{center}
\caption{Schematic illustration of the clustering-validity criterion. Each dot represents a student solution to a problem or problem part; numbers and colors represent human-assigned reference scores. Clusters are formed without using score information. If embedding clusters supported grading-related interpretation, sizable clusters would contain responses with similar scores. Mixed-score clusters may still be useful for exploratory organization, whereas clusters with large score variation indicate that embedding proximity does not preserve the grading-relevant construct. Single-member clusters may represent unusual solution approaches or failed grouping.}
\label{fig:clusters}
\end{figure}

\subsection{Testing embedding-based similarity and clustering}
The central premise tested in this study is that proximity in embedding space may correspond to similarity in human-assigned scores --- ideally, clusters would look like those depicted in Fig.~\ref{fig:clusters}. We evaluated this premise in two complementary ways. First, we asked whether pairwise embedding similarity predicts pairwise score similarity: for each pair of student solutions to the same problem, we compared the embedding similarity of the two responses with the negative absolute difference between their human-assigned scores. Second, we asked whether hierarchical clusters formed from embedding similarities produced score-homogeneous groups. In both analyses, human scores were used only as an external reference standard for evaluation and were not used during transcription, embedding generation, similarity computation, or clustering.

This framing treats clustering as an object of validation rather than as an assumed grading method. Embedding-based clusters may be useful for organizing responses, selecting examples, or supporting human review, but a grading-related interpretation requires evidence that cluster membership preserves rubric-relevant distinctions.

\subsection{Research questions.}
The main hypothesis of this study is
\begin{quote}
If embeddings are meaningful semantic-conceptual representations of student-problem solutions, then proximity in embedding space should correspond to similarity in human-assigned correctness scores.
\end{quote}
This study addresses four research questions.
\begin{enumerate}
\item To what extent does pairwise similarity in embedding space predict similarity in human-assigned scores for handwritten physics solutions?
\item Do hierarchical clusters formed from embedding similarities produce groups that are homogeneous with respect to human-assigned scores?
\item How do these results depend on the textual representation of the handwritten work, the embedding mechanism, the similarity metric, and the clustering method?
\item Are the observed relationships strong enough to justify grading-related interpretations of embedding clusters, or are the clusters better understood as exploratory groupings requiring human validation?
\end{enumerate}

\section{Setting}

\subsection{Institution}
ETH Zurich is a technical university enrolling approximately 25{,}000 students from more than 120 countries. Admission is highly selective for international applicants, while Swiss high school graduates have open access. Most undergraduate instruction is in German. In line with the academic tradition of German-speaking universities, assessment is primarily summative, with high-stakes examinations at the end of courses rather than continuous assessment during the semester.

\begin{figure}
\begin{center}
\includegraphics[width=\columnwidth]{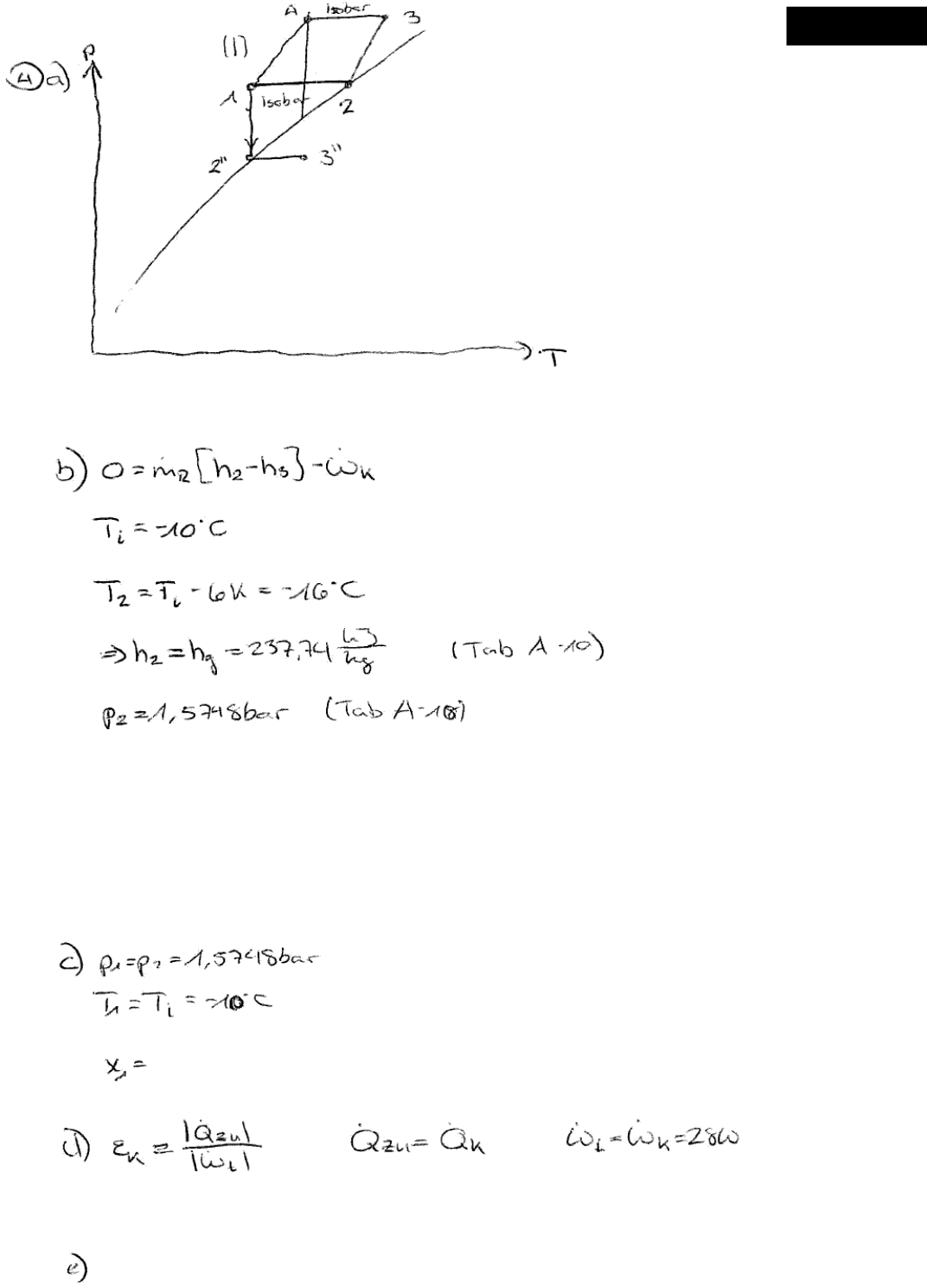}
\end{center}
\caption{Example of student work on the exam. The black box in the top-right is the redacted student name.}
\label{fig:example}
\end{figure}

\subsection{Exam}
We analyze a high-stakes exam in engineering thermodynamics~\cite{kortemeyer2024grading,kortemeyer2025assessing} covering standard topics such as energy, exergy, entropy, and enthalpy. The exam had four problems, 1~to~4, each with a handful of problem parts (a~to~d or a~to~e), and those in turn were graded on a handful of rubric items each (85 rubric items in total). Students produced complete handwritten solutions using permanent pens and were instructed to leave out any work they did not wish to be graded. Figure~\ref{fig:example} shows an example of student work on this exam.

\subsection{Sample}
The study was approved by ETH Zurich Ethics Committee (Study~2023-N-286). Of the 434~students who sat the exam, 252~provided informed consent and were included in the research dataset. Consent forms were distributed before the exam and collected with the submissions. Because participation was opt-in in a high-stakes context, selection effects are possible, for example, students may have ignored the form under time pressure or withheld consent after perceiving poor performance. The research protocol did not permit linking participation status to demographics or scores, so direct comparisons between participants and non-participants are not available. The resulting dataset comprises 3041~pages of handwritten solutions (mean 12~pages per student) spanning a wide range of performance; as not all students worked on all problems, the dataset resulted in 993~student-problem solutions (instead of~1008).

\section{Methods}

\subsection{Transcription of Handwriting}\label{sec:trans}

The scanned handwritten solutions were converted into text under five transcription conditions using GPT-5.4~\cite{gpt54}. 
These conditions were designed to vary the degree of normalization and interpretation applied to the handwritten artifact while keeping the task non-evaluative. 
In all five conditions, the multimodal language model was instructed to describe only what was visibly present in the student work. 
It was explicitly instructed not to solve the problem, correct the solution, assign points, identify errors as errors, or compare the work to an official or ideal solution. 
The model was also instructed to preserve visible variables, units, state labels, numerical values, final answers, crossed-out work when readable, and uncertainty about illegible or ambiguous content. 
Thus, the transcription conditions differed in representational form, but not in their prohibition against grading or correction.

\begin{figure*}'
\includegraphics[width=0.7\textwidth]{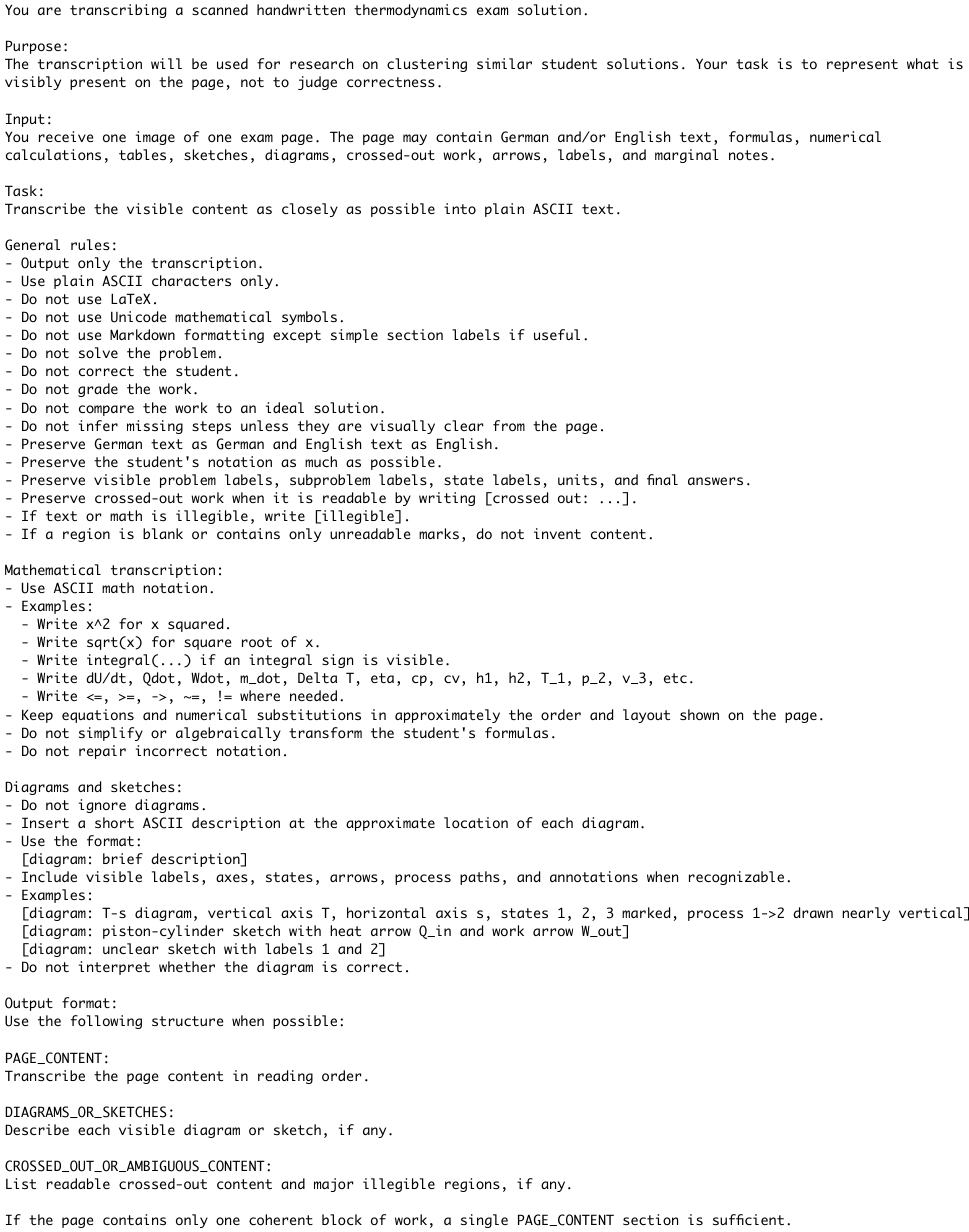}
\caption{Prompt for mode \texttt{ascii}.}
\label{fig:asciipr}
\end{figure*}

\begin{figure*}'
\includegraphics[width=0.7\textwidth]{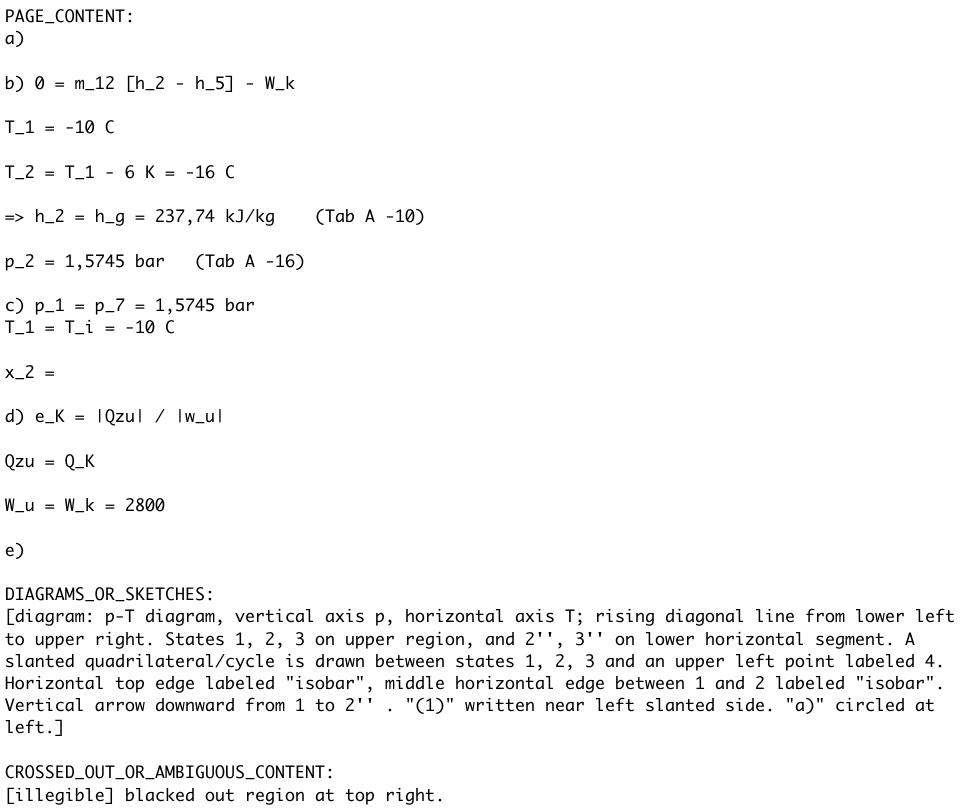}
\caption{Transcription in mode \texttt{ascii} of the student work in Fig.~\ref{fig:example}.}
\label{fig:ascii}
\end{figure*}

\begin{figure*}'
\includegraphics[width=0.7\textwidth]{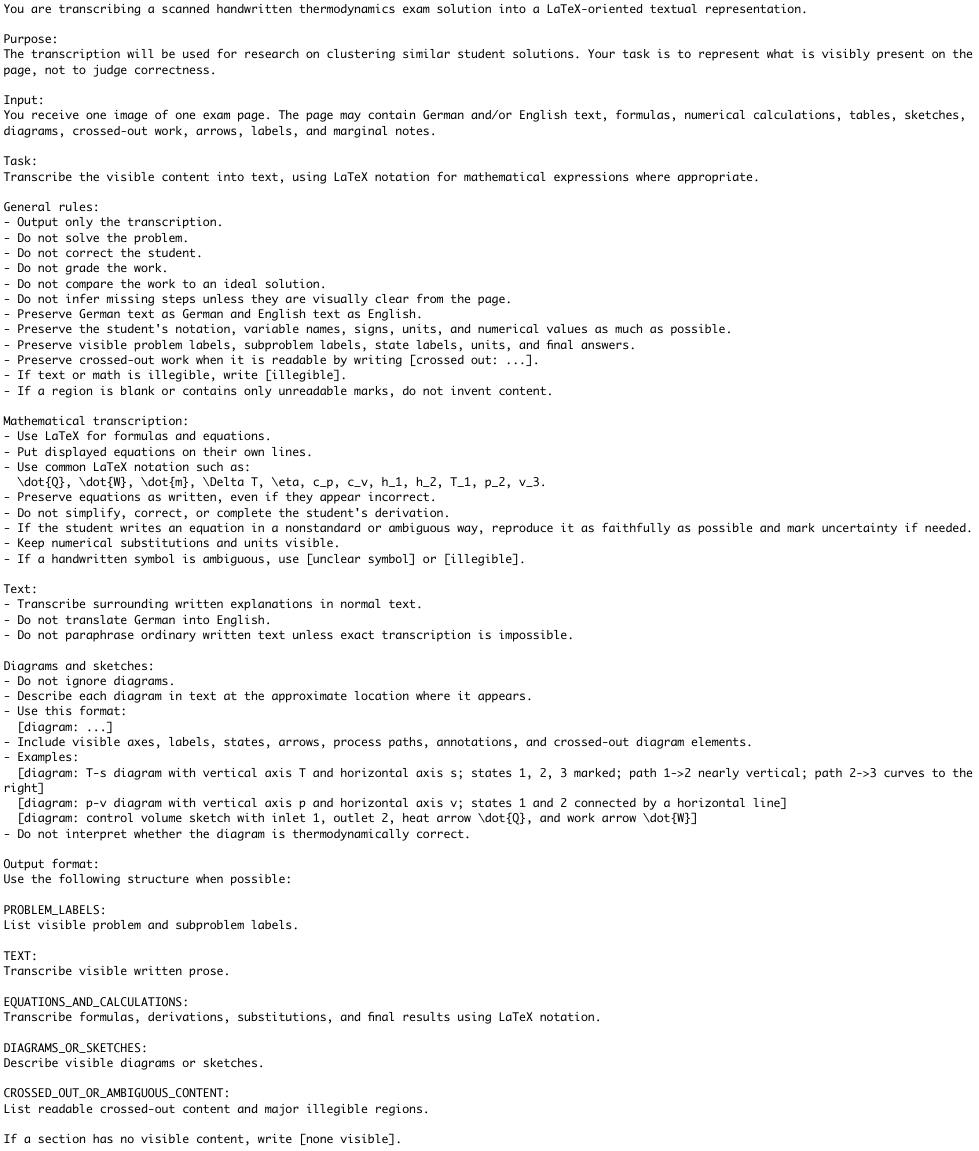}
\caption{Prompt for mode \texttt{latex}.}
\label{fig:latexpr}
\end{figure*}

\begin{figure*}
\includegraphics[width=0.6\textwidth]{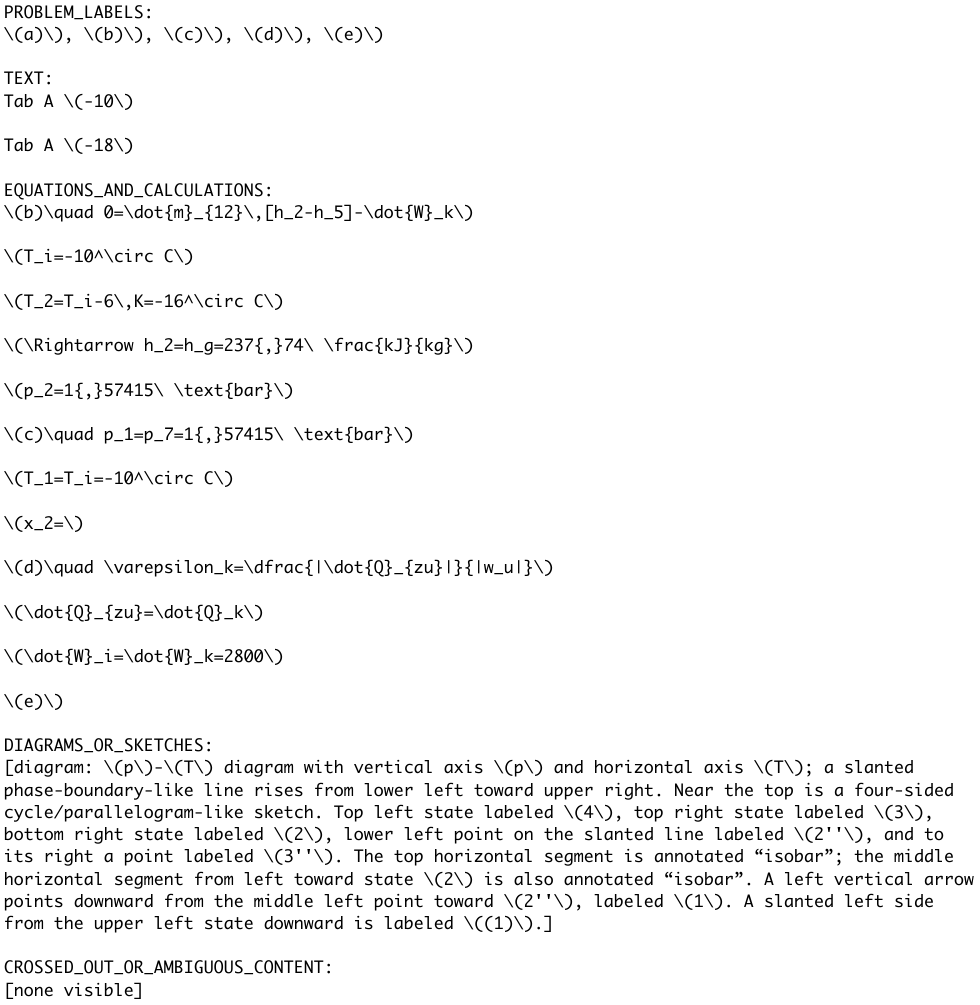}
\caption{Transcription in mode \texttt{latex} of the student work in Fig.~\ref{fig:example}.}
\label{fig:latex}
\end{figure*}

\begin{figure*}'
\includegraphics[width=0.8\textwidth]{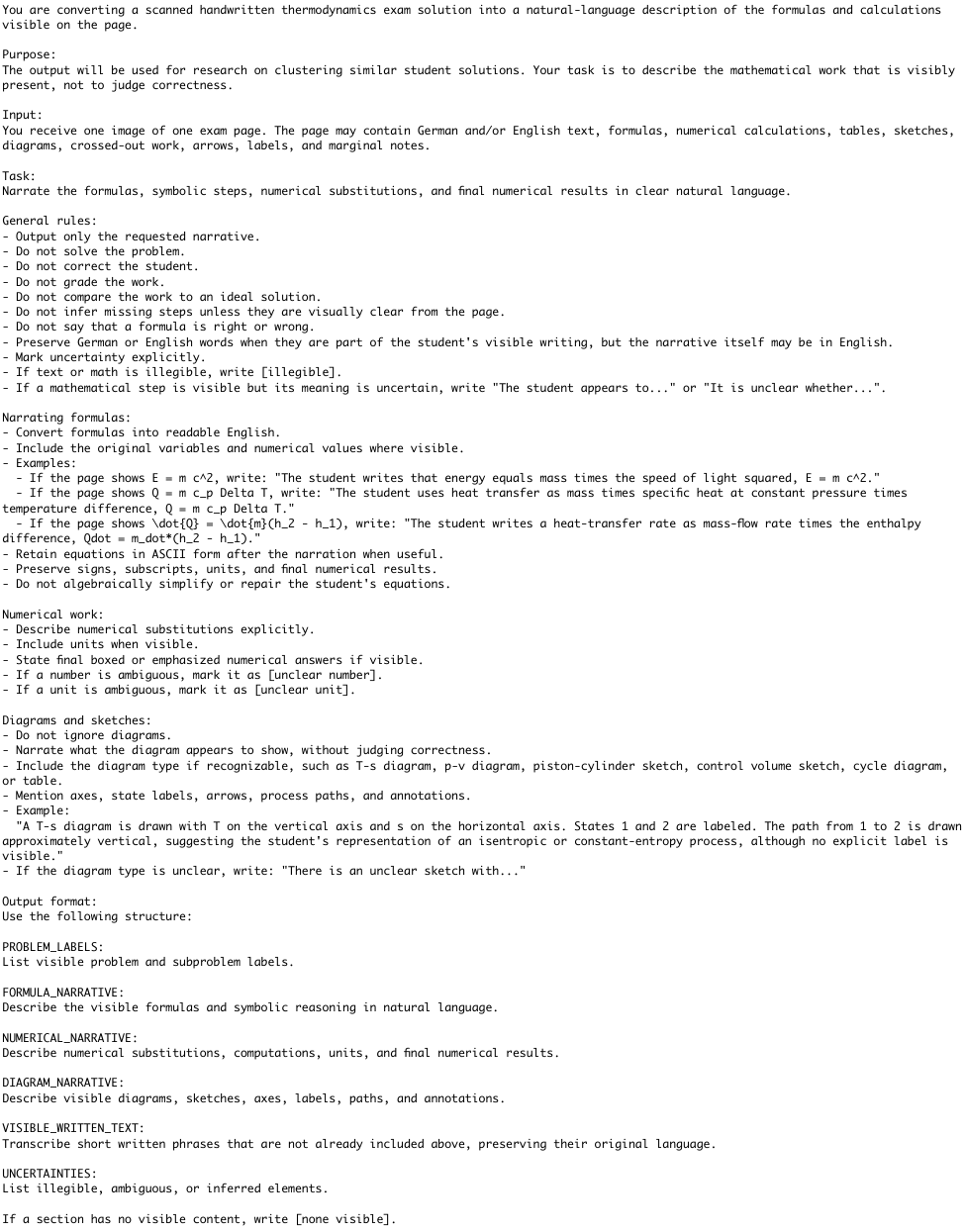}
\caption{Prompt for mode \texttt{formula\_narrative}.}
\label{fig:formulanarrativepr}
\end{figure*}

\begin{figure*}
\includegraphics[width=0.9\textwidth]{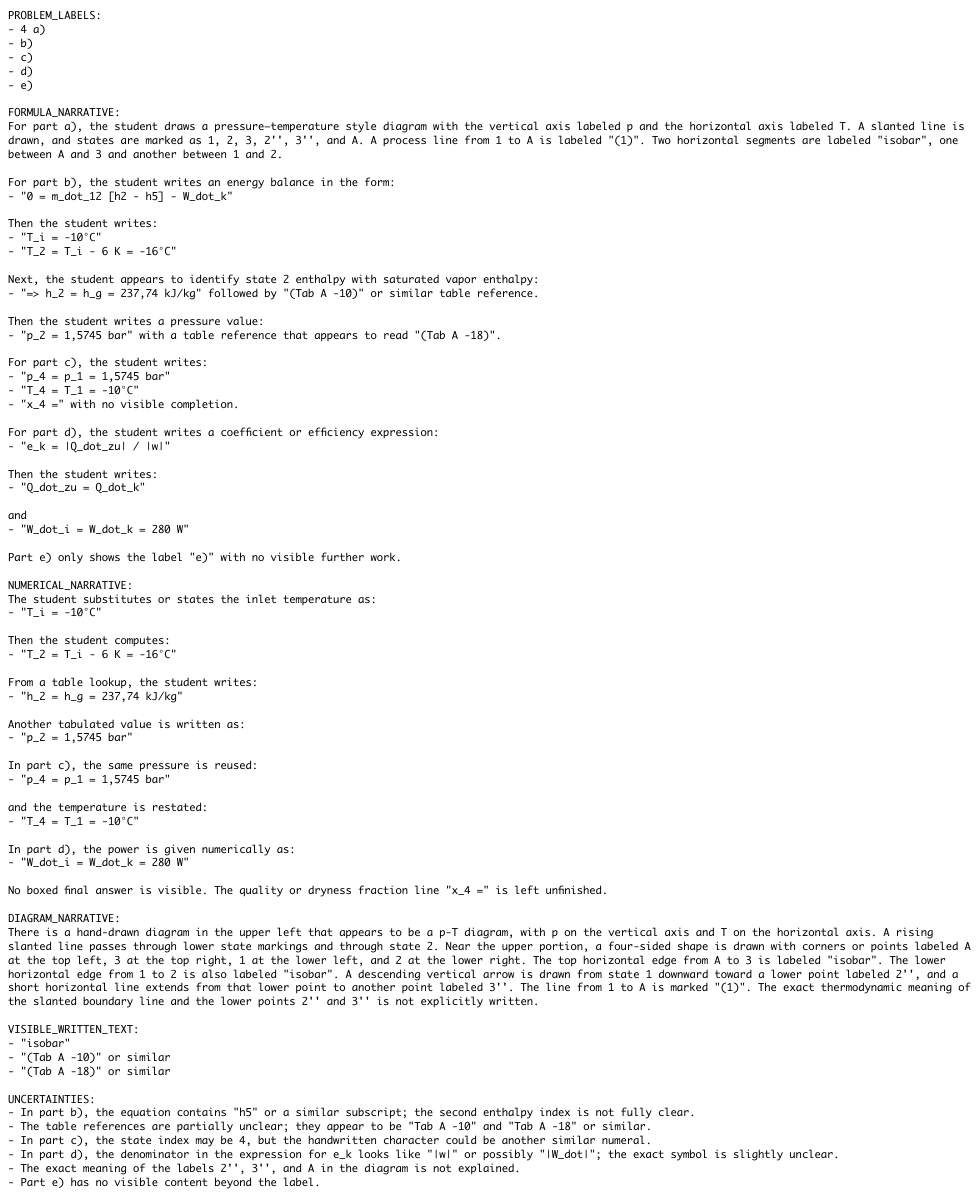}
\caption{Transcription in mode \texttt{formula\_narrative} of the student work in Fig.~\ref{fig:example}.}
\label{fig:formulanarrative}
\end{figure*}

\begin{figure*}'
\includegraphics[width=0.8\textwidth]{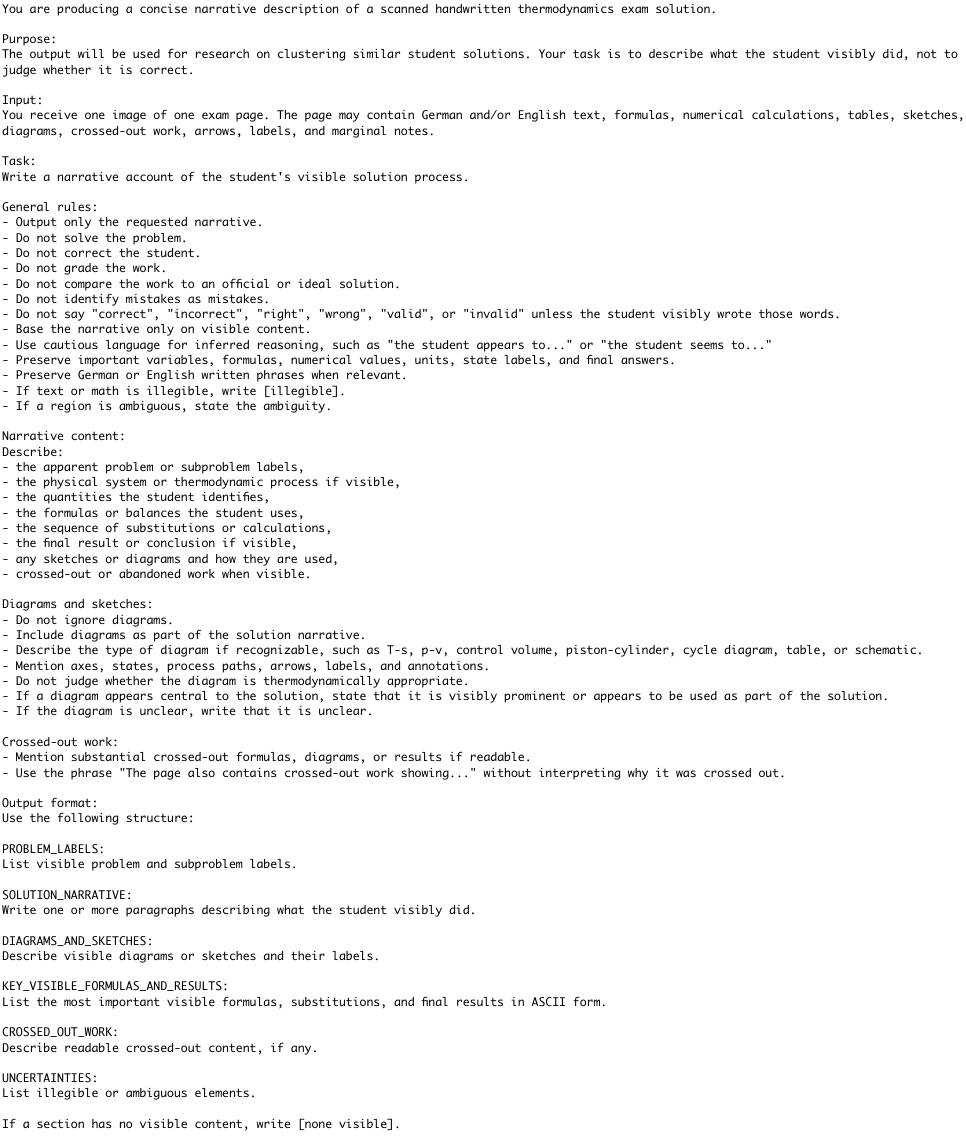}
\caption{Prompt for mode \texttt{solution\_narrative}.}
\label{fig:solutionnarrativepr}
\end{figure*}

\begin{figure*}
\includegraphics[width=0.9\textwidth]{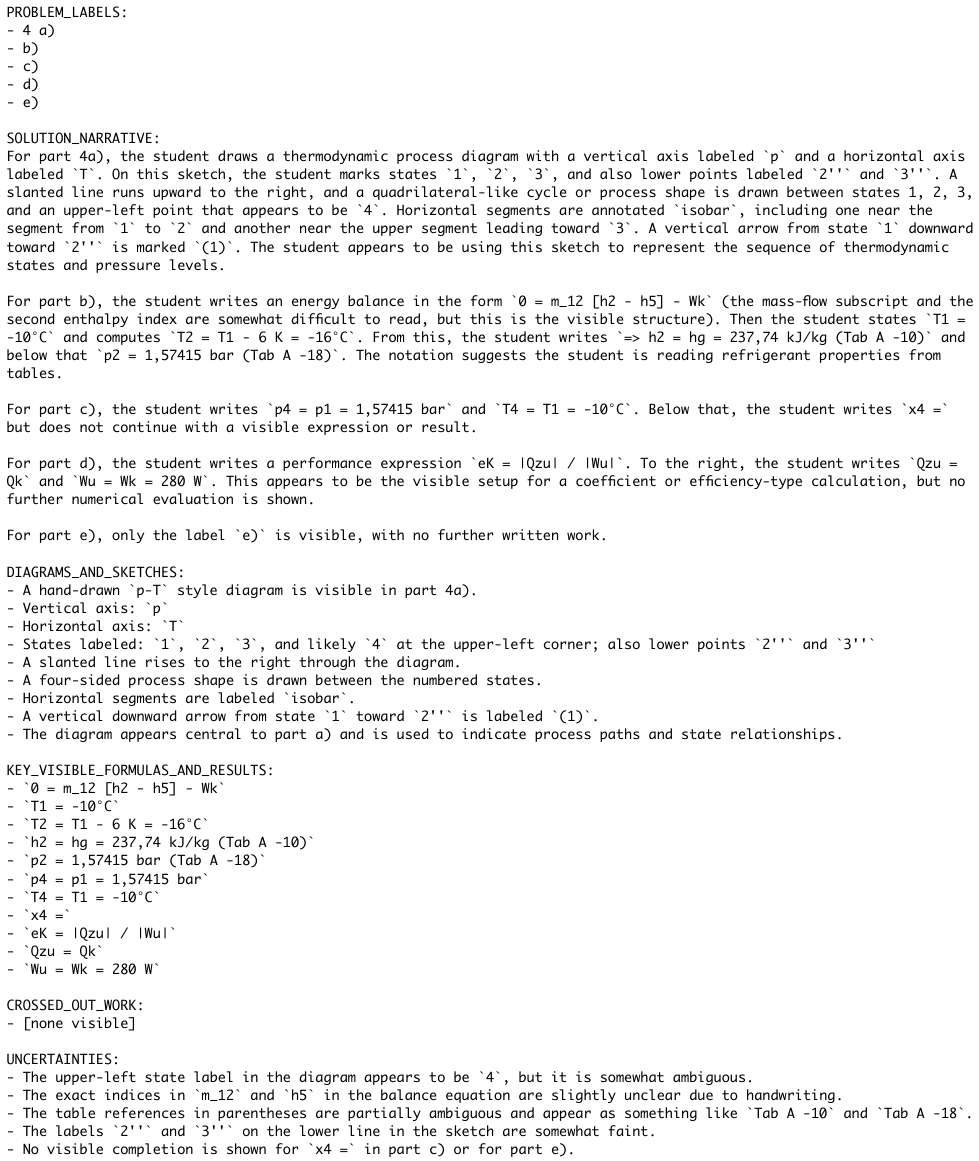}
\caption{Transcription in mode \texttt{solution\_narrative} of the student work in Fig.~\ref{fig:example}.}
\label{fig:solutionnarrative}
\end{figure*}

\begin{figure*}'
\includegraphics[width=0.8\textwidth]{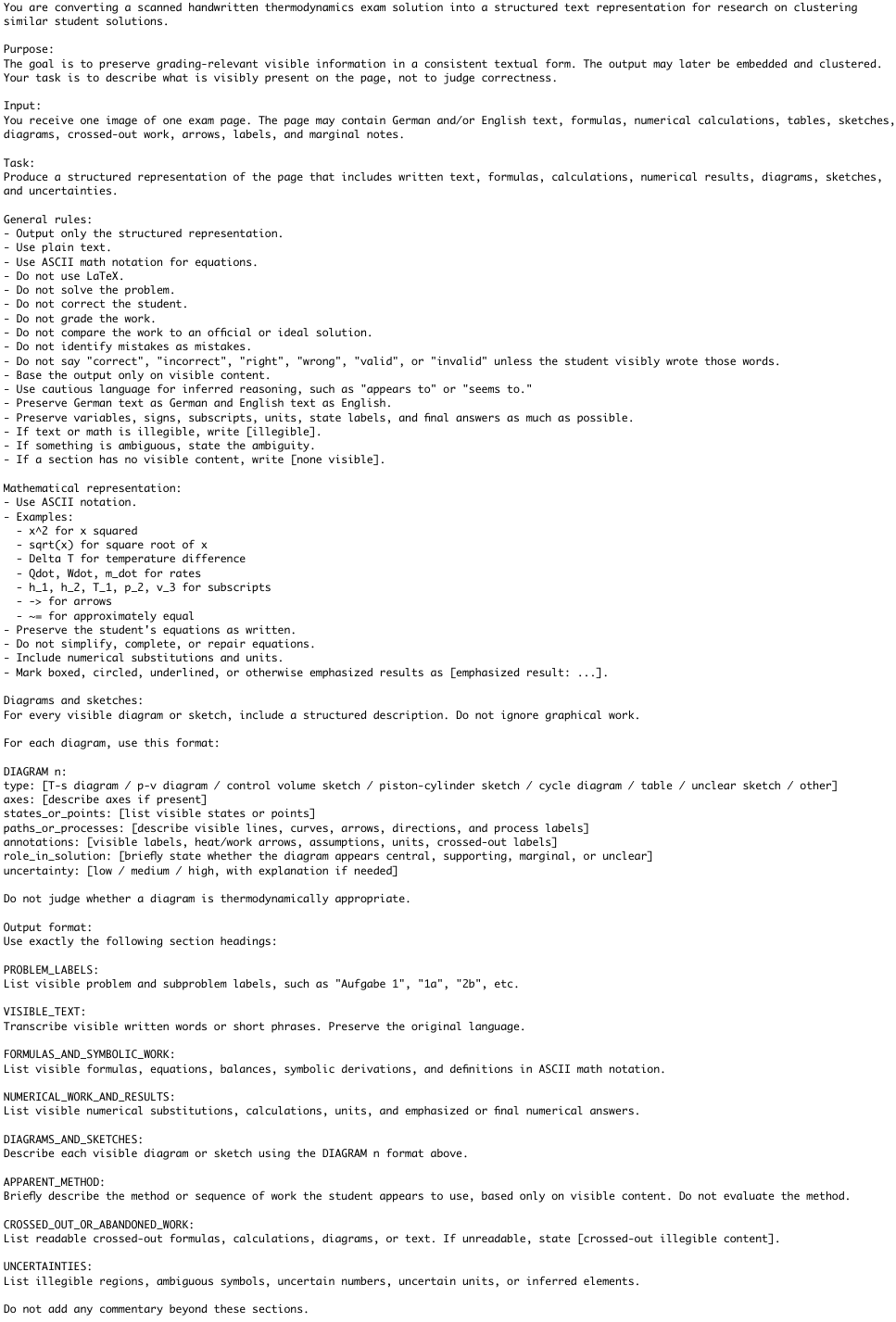}
\caption{Prompt for mode \texttt{hybrid\_structured}.}
\label{fig:hybridstructuredpr}
\end{figure*}

The five transcription conditions were as follows.
\begin{description}
\item[ASCII verbatim (\texttt{ascii})]
In this condition, the model was asked to transcribe the visible content as closely as possible into plain ASCII text (see Fig.~\ref{fig:asciipr} for the prompt). 
Mathematical notation was represented using ASCII conventions, such as \texttt{x\^{}2}, \texttt{sqrt(x)}, \texttt{Delta T}, \texttt{Qdot}, \texttt{m\_dot}, \texttt{h\_1}, and \texttt{p\_2}. 
Diagrams and sketches were not ignored, but were represented only by brief textual descriptions placed near their approximate location in the page content (see Fig.~\ref{fig:ascii} for an example).

The rationale for this condition was to produce a minimally normalized representation of the original student artifact. 
It preserves idiosyncratic notation, ordering, partial work, and visible structure as much as possible. 
This condition therefore serves as a relatively low-interpretation baseline. 
However, because it retains substantial surface variation in notation, layout, and handwriting-derived transcription noise, it may be less effective for clustering solutions that are mathematically or conceptually similar but written in different forms.

\item[LaTeX-oriented (\texttt{latex})]
In this condition, the model was asked to transcribe visible mathematical expressions using LaTeX notation while preserving surrounding prose in ordinary text (Fig.~\ref{fig:latexpr}). 
For example, rates, subscripts, and thermodynamic quantities could be rendered as \verb|\dot{Q}|, \verb|\dot{m}|, \verb|\Delta T|, \verb|h_1|, or \verb|c_p|. 
The instruction emphasized that equations should be preserved as written, even if they appeared nonstandard or incomplete, and that the model should not simplify, repair, or complete derivations. 
Diagrams were described textually, including axes, labels, states, arrows, and process paths where visible (Fig.~\ref{fig:latex}).

The rationale for this condition was to obtain a more standardized representation of mathematical work than the ASCII condition while retaining the symbolic structure of the student's solution. 
This representation may be useful for calculation-heavy or derivation-heavy problems, where similarity may depend on the equations selected, the variables related, and the substitutions performed. 
At the same time, LaTeX syntax may introduce tokens that are not semantically meaningful to all embedding models, and the act of converting handwriting into LaTeX may normalize away some student-specific variation. 
Thus, this condition tests whether mathematical standardization improves clustering relative to a more literal transcription.

\item[Formula narrative (\texttt{formula\_narrative})]
In this condition, the model was asked to narrate the formulas, symbolic steps, numerical substitutions, and final numerical results in natural language (Fig.~\ref{fig:formulanarrativepr}). 
For example, an expression such as \texttt{Q = m c\_p Delta T} would be represented as a statement that the student uses heat transfer as mass times specific heat at constant pressure times temperature difference, while retaining the original formula and visible quantities where useful. 
The narrative was constrained to describe visible work only; uncertain inferences were to be marked with phrases such as ``the student appears to'' or ``it is unclear whether'' (Fig.~\ref{fig:formulanarrative}).

The rationale for this condition was to make mathematical content more accessible to general-purpose text embeddings. 
Students may write mathematically equivalent or closely related expressions using different notation, ordering, or intermediate variables. 
A formula narrative may reduce superficial notational differences by converting symbolic work into semantically similar natural-language descriptions. 
This condition is therefore expected to be useful when the important similarity among solutions lies in the underlying formula choices, substitutions, or computational strategy rather than in the exact notation used. 
Its potential disadvantage is that narration introduces a greater degree of model mediation and may lose fine-grained algebraic or numerical detail.

\item[Solution narrative (\texttt{solution\_narrative})]
In this condition, the model was asked to produce a concise narrative account of what the student visibly did in the solution (Fig.~\ref{fig:solutionnarrativepr}). 
The output described the apparent problem labels, physical system or process when visible, quantities identified, formulas used, sequence of substitutions or calculations, final results, and any diagrams or crossed-out work. 
The model was instructed to use cautious language for inferred reasoning and to avoid evaluative terms such as ``correct,'' ``incorrect,'' ``right,'' or ``wrong'' unless such words were visibly written by the student (Fig.~\ref{fig:solutionnarrative}).

The rationale for this condition was to represent the student's apparent solution process rather than only the literal written content or isolated formulas. 
This may be especially useful for conceptual problems, problems involving multi-step reasoning, and problems where a diagram or choice of method is central to the response. 
For clustering, the hypothesis is that students who follow similar reasoning paths or make similar methodological choices may be embedded near one another even when their exact notation or numerical work differs. 
The risk is that this condition is the most interpretive of the five: the model may impose a coherent narrative on incomplete or ambiguous work. 
For this reason, the prompt emphasized visible evidence and explicit uncertainty.

\item[Hybrid structured (\texttt{hybrid\_structured})]
In this condition, the model was asked to produce a structured representation with fixed section headings: problem labels, visible text, formulas and symbolic work, numerical work and results, diagrams and sketches, apparent method, crossed-out or abandoned work, and uncertainties (Fig.~\ref{fig:hybridstructuredpr}). 
Mathematical expressions were represented in ASCII notation. 
For each visible diagram or sketch, the model was asked to provide a structured description including the diagram type, axes, states or points, paths or processes, annotations, apparent role in the solution, and uncertainty. 
As in the other conditions, the model was instructed not to evaluate correctness  (Fig.~\ref{fig:hybridstructured}). 

The rationale for this condition was to combine the strengths of the previous conditions. 
It preserves literal written content, symbolic and numerical work, diagrammatic information, and a brief non-evaluative description of the apparent method, while imposing a consistent structure across solutions. 
This condition was designed specifically as an embedding-oriented representation: stable section headings provide a common schema, while the contents preserve solution-specific information likely to be relevant for clustering. 
It may be particularly valuable for thermodynamics exams, where grading-relevant information can appear in prose, equations, numerical substitutions, property-table lookups, cycle sketches, control-volume diagrams, or thermodynamic state diagrams. 
The hybrid condition therefore tests whether a structured, multimodal textual representation yields clusters that better align with human-assigned scores than either literal transcription or free narrative alone.
\end{description}

\begin{figure*}
\includegraphics[width=\textwidth]{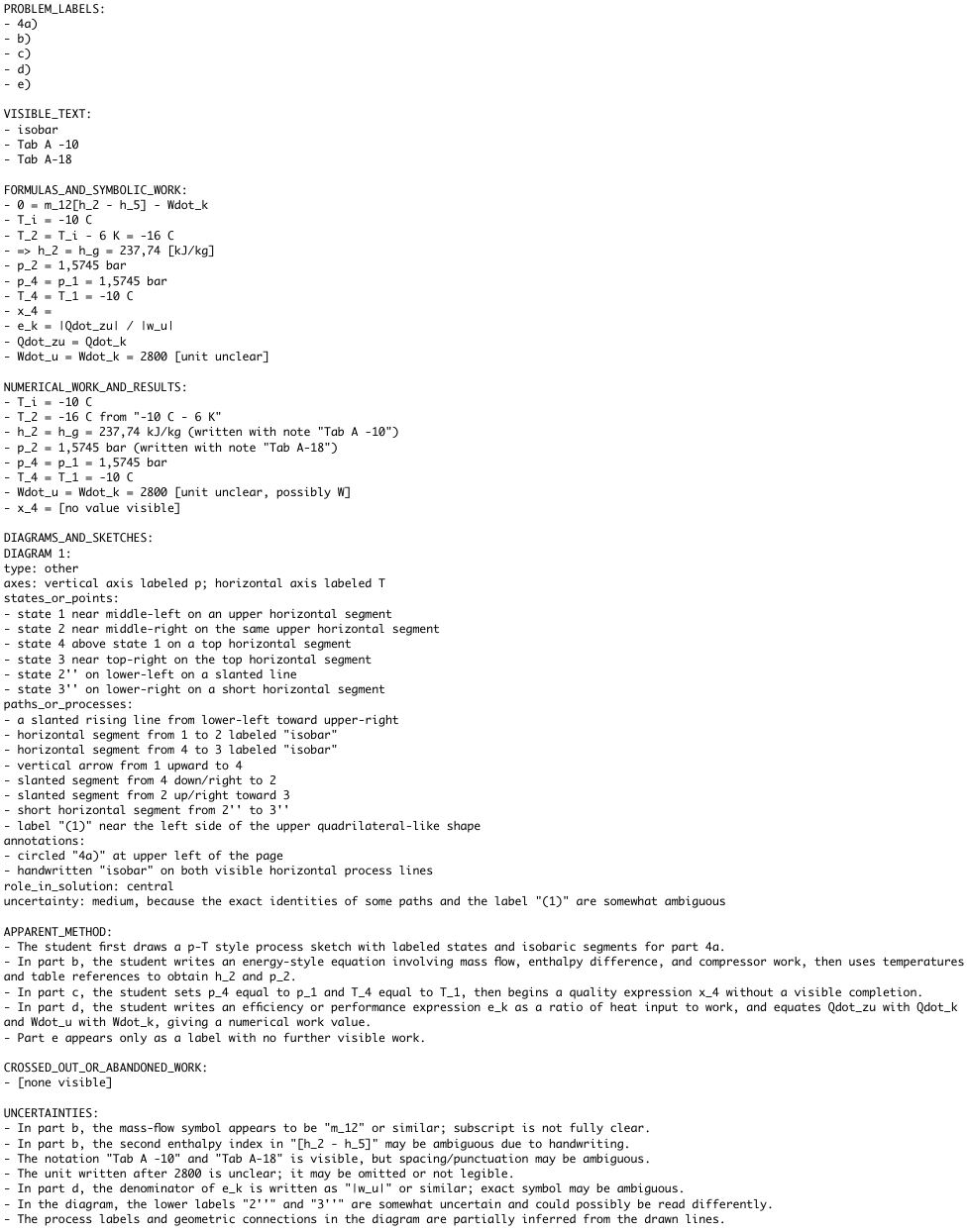}
\caption{Transcription in mode \texttt{hybrid\_structured} of the student work in Fig.~\ref{fig:example}.}
\label{fig:hybridstructured}
\end{figure*}

\subsection{Common constraints across transcription conditions}
\label{subsec:transcription-common-constraints}

All transcription conditions shared the following constraints:
\begin{enumerate}
\item the model operated only on the visible student artifact. 
It was not given the official solution, grading rubric, or point score at transcription time. 
\item the task was framed as transcription or description, not assessment. 
The model was explicitly instructed not to solve, correct, grade, or compare the student work to an ideal solution. 
\item ambiguity was to be preserved rather than resolved: illegible material was marked as such, uncertain symbols or numbers were identified as uncertain, and inferred reasoning was to be described cautiously. 
\item diagrams and sketches were included in the textual representation rather than discarded. 
This was important because thermodynamics solutions may contain grading-relevant graphical elements, such as $T$--$s$ diagrams, $p$--$v$ diagrams, cycle sketches, process paths, state labels, control-volume schematics, and heat or work arrows. 
\item crossed-out or abandoned work was retained when readable, since such work may reflect solution strategies or misconceptions relevant to similarity clustering.
\end{enumerate}

The purpose of comparing these five transcription conditions was not to determine which one most faithfully reproduces the handwriting in a human sense, but rather to determine which textual representation best preserves grading-relevant similarity among student solutions. 
The subsequent clustering analysis therefore treats the transcription condition as an experimental variable. 
For each condition, embeddings are computed from the resulting text, clusters are formed within each problem, and the distribution of human-assigned scores within each cluster is examined. 
A useful transcription condition is one for which clusters contain solutions that are similar not only in embedding space, but also in the scores assigned by human graders and, upon inspection, in their apparent solution strategies or recurring errors.

\subsection{Text-embedding methods}
We generated vector representations of each student response using nine embedding mechanisms. In the following, the short names in parentheses are used throughout the remainder of the paper.

\begin{description}

\item[OpenAI text-embedding-3-large (\texttt{oai3large})]
This condition used OpenAI's \texttt{text-embedding-3-large} embedding model as a hosted, general-purpose text-embedding baseline. The model maps each input response representation to a dense vector intended to support semantic comparison of text by vector similarity. In this study, \texttt{oai3large} serves as a strong proprietary reference embedding that was not specialized to physics education or to the particular scoring rubrics used here~\cite{openai_text_embedding_3_large}.

\item[Jina text-retrieval F16 (\texttt{jina4retF16})]
This condition used the Jina Embeddings v4 model with the text-retrieval adapter in F16 GGUF format. Jina Embeddings v4 is a multimodal, multilingual embedding model with task-specific adapters. The text-retrieval adapter is optimized for retrieval-style use cases, where vector similarity is intended to support finding relevant passages or documents. We used it here as an embedding mechanism for student responses, allowing us to test whether retrieval-optimized semantic structure is aligned with grading-relevant similarity~\cite{jina_embeddings_v4,gunther_etal_2025_jina_embeddings_v4}.

\item[Jina text-retrieval Q8 (\texttt{jina4retQ8})]
The same adapter as \texttt{jina4retF16}, but in an 8-bit quantized GGUF representation. This condition was included to test whether reduced-precision local inference preserved the grading-relevant geometry observed with the higher-precision retrieval-adapter condition.

\item[Jina text-retrieval Q4\_K\_M (\texttt{jina4retQ4})]
The same adapter as \texttt{jina4retF16}, but in a more aggressively quantized Q4\_K\_M GGUF representation. It was included as a low-resource local embedding condition, testing whether a substantially compressed retrieval-oriented model still preserved enough semantic structure to be useful for clustering or score-similarity analysis.

\item[Jina text-matching F16 (\texttt{jina4matchF16})]
This condition used Jina Embeddings v4 with the text-matching adapter in F16 GGUF format. Unlike the retrieval adapter, the text-matching adapter is intended for symmetric semantic-similarity tasks such as duplicate detection, clustering, and matching texts of the same general type. This made it a particularly relevant comparison condition for the present study, since student responses are being compared to other student responses rather than to external query documents \citep{jina_embeddings_v4,gunther_etal_2025_jina_embeddings_v4}.

\item[Jina text-matching Q8 (\texttt{jina4matchQ8})]
This condition tests whether the symmetric text-matching geometry of \texttt{jina4matchF16} is retained under moderate 8-bit quantization.

\item[Jina text-matching Q4\_K\_M (\texttt{jina4matchQ4})]
This condition represents the most compressed version of the Jina text-matching family used in the study, and was included to assess the robustness of embedding-based clustering and score-similarity analyses under low-resource local deployment.

\item[Qwen 2.5-Math-1.5B-Instruct (\texttt{qwenMath1p5B})]
This condition used the instruction-tuned 1.5B-parameter Qwen2.5-Math model as a math-specialized representation mechanism. Unlike the dedicated embedding models above, Qwen2.5-Math-Instruct is primarily a generative mathematical reasoning model rather than a conventional embedding model. We thus extracted final-layer hidden-state representations and converted them to one vector per response by mean pooling over non-padding tokens. The resulting vectors were not normalized before similarity computation. We included it to test whether representations derived from a smaller mathematics-specialized instruction model better preserved grading-relevant distinctions in physics solutions than general-purpose text embeddings \citep{qwen25math}.

\item[Qwen 2.5-Math-7B-Instruct (\texttt{qwenMath7B})]
This condition used the instruction-tuned 7B-parameter Qwen2.5-Math model as a larger math-specialized representation mechanism. As with \texttt{qwenMath1p5B}, this model was not included as a standard sentence-embedding model, but as a test of whether a mathematics-specialized language model produced response representations whose geometry was more aligned with rubric-relevant correctness. This condition also allowed us to examine whether increased model scale within the Qwen2.5-Math family improved the usefulness of embedding-like representations for clustering and grading-related similarity analyses \citep{qwen25math}.

\end{description}

For each of the nine embedding mechanisms, we generated embeddings for five response representations: \texttt{ascii}, \texttt{latex}, \texttt{formula\_narrative}, \texttt{solution\_narrative}, and \texttt{hybrid\_structured}. The four problems contained 245, 247, 252, and 249 student responses, respectively, for a total of 993 student-problem responses per response representation. Thus, the full design called for
$
9 \times 5 \times 993 = 44{,}685
$
embedding vectors. During downstream analysis, after the GPUs used for local embedding generation were no longer available, we discovered one corrupted embedding file, corresponding to student response for Problem~4 in one embedding condition. Because this vector could not be regenerated at that stage, we excluded the corresponding student-problem case globally rather than allowing different embedding mechanisms to be evaluated on different response sets. Specifically, the student-problem was removed from consideration for all embedding mechanisms and all response representations. The final harmonized analysis set therefore contained 245, 247, 252, and 248 responses for Problems 1--4, respectively, or 992 student-problem responses per response representation. This yielded
$
9 \times 5 \times 992 = 44{,}640
$
usable embedding vectors in the final comparative analyses.

\subsection{Similarity and clustering evaluation}

For each problem, transcription condition, and embedding mechanism, we constructed a matrix of pairwise similarities among student-problem responses. We evaluated these matrices in two complementary ways:
\begin{itemize}
\item First, we asked whether pairwise embedding similarity predicted pairwise score similarity.
\item Second, we asked whether hierarchical clusters formed from embedding similarities were homogeneous with respect to human-assigned scores.
\end{itemize}
In both analyses, human scores were used only as an external reference standard for evaluation and were not used during transcription, embedding generation, similarity computation, or clustering.
\subsubsection{Similarity}
Let $x_i \in \mathbb{R}^d$ denote the embedding vector for student response $i$ within a fixed problem, transcription condition, and embedding mechanism. We computed two similarity matrices. The first was ordinary cosine similarity,
\begin{equation}\label{eq:cos}
s_{ij}^{\mathrm{cos}} =
\frac{x_i \cdot x_j}{\lVert x_i\rVert \lVert x_j\rVert}.
\end{equation}
The second was mean-centered cosine similarity. For each embedding matrix (fixed problem, transcription condition and embedding mechanism), we computed the mean vector
\begin{equation}
\mu = \frac{1}{n}\sum_{i=1}^{n}x_i
\end{equation}
and then compared centered vectors $z_i=x_i-\mu$:
\begin{equation}\label{eq:centercos}
s_{ij}^{\mathrm{cent}} =
\frac{z_i \cdot z_j}{\lVert z_i\rVert \lVert z_j\rVert}.
\end{equation}

The centered measure was included because all responses to a given problem share substantial semantic context from the problem statement and the domain. Mean-centering removes the common direction of the response set and may therefore emphasize distinctions among student solutions rather than similarity arising merely from answering the same problem. Similarity matrices were checked for finite entries, symmetry, diagonal values close to one, and values within the numerical range expected for cosine-like similarities.

For the pairwise score-similarity analysis, we considered all unordered pairs of student responses $(i,j)$ within the same problem for which both responses had human-assigned scores. For each pair, we recorded the embedding similarity $s_{ij}$, the human scores $y_i$ and $y_j$, the absolute score difference
\begin{equation}\label{eq:delta}
\Delta_{ij}=|y_i-y_j|,
\end{equation}
and the negative absolute score difference $-\Delta_{ij}$. If embedding similarity preserved grading-relevant distinctions, higher values of $s_{ij}$ should correspond to smaller score differences, and therefore to larger values of $-\Delta_{ij}$. We therefore computed Pearson and Spearman correlations between $s_{ij}$ and $-\Delta_{ij}$. Spearman correlation, rather than Pearson correlation, was treated as the primary pairwise quality measure, because it tests whether the embedding similarity orders pairs in a grading-relevant way without assuming a linear relationship between similarity values and score differences. Pearson correlation was retained as a secondary measure. We also binned pairs by similarity quantile and computed the mean and median absolute score difference within each bin as a diagnostic visualization of whether high-similarity pairs had smaller score differences.

Because each student response appears in many pairs, the pairwise observations are not statistically independent. We therefore treat the Pearson and Spearman correlations as descriptive effect-size measures of alignment between embedding similarity and score similarity, rather than as the basis for significance testing over independent observations.

\subsubsection{Clustering}
For the clustering analysis, each similarity matrix $S$ was converted to a distance matrix
\begin{equation}\label{eq:dij}
D_{ij}=1-S_{ij}.
\end{equation}
We then performed agglomerative hierarchical clustering using average linkage and complete linkage. These two linkage methods represent different assumptions about what should count as a coherent group: average linkage merges clusters based on average inter-cluster distance, whereas complete linkage is more conservative and penalizes clusters whose most distant members remain far apart. For each linkage method, we evaluated clusterings with requested numbers of clusters $k=2,\ldots,80$, capped at $n-1$ for a response set of size $n$. Cluster assignments were produced without using score information.

For each clustering, we summarized the distribution of human scores within each cluster. For a cluster $c$, with score set ${y_i : i \in c}$, we computed the cluster size $n_c$, mean score, score standard deviation, score range, and within-cluster sum of squares,
\begin{equation}
\mathrm{WSS}_c = \sum_{i \in c}(y_i-\bar{y}_c)^2\ .
\end{equation}
Across all clusters, we computed the total within-cluster sum of squares
\begin{equation}
\mathrm{WSS}=\sum_{c=1}^{K} \mathrm{WSS}_c\ ,
\end{equation}
where $K$ is the number of clusters, and compared it with the total score sum of squares,
\begin{equation}
\mathrm{TSS}=\sum_i(y_i-\bar{y})^2.
\end{equation}

This yielded an $R^2$-like measure,
\begin{equation}
R^2_{\mathrm{clusters}} = 1-\frac{\mathrm{WSS}}{\mathrm{TSS}}\ ,
\end{equation}
which describes how much of the score variance is accounted for by cluster membership.

Because very small clusters are of limited practical value for grading or batching workflows, clusters with fewer than five students were not counted as useful clusters in the coverage analysis. We therefore computed useful-cluster coverage, defined as the fraction of students assigned to clusters of size at least five. For the students in useful clusters, we computed the weighted mean cluster score standard deviation,
\begin{equation}\label{eq:sigmauseful}
\bar{\sigma}_{\mathrm{useful}} =
\frac{\sum_{c:n_c\ge 5} n_c\sigma_c}
{\sum_{c:n_c\ge 5} n_c}\ ,
\end{equation}
and the corresponding weighted mean score range. Lower values of these quantities indicate more score-homogeneous clusters, while higher useful-cluster coverage indicates that the method organizes a larger fraction of the response set into nontrivial groups.
The upper limit of cluster numbers $k=80$ and the minimum useful cluster size of five were chosen to allow relatively fine-grained grouping while excluding singleton or very small clusters that would not meaningfully reduce grading or review effort.

For aggregate comparisons across problems, we treated a clustering configuration as potentially useful only if at least 70\% of students were assigned to useful clusters and at least two useful clusters were present. Within each problem and experimental condition (transcription and embedding method), we then selected the value of $k$ with the lowest weighted mean score standard deviation among useful clusters. These best-per-problem rows were aggregated across the four problems by transcription condition, embedding mechanism, similarity method, and linkage method. This produced the overall clustering-quality summaries reported in the Results section. The selection of $k$ was used only for retrospective evaluation of the best observed clustering quality for each condition. It should therefore be interpreted as an optimistic diagnostic of the embedding geometry, not as a deployable unsupervised procedure for choosing $k$ without access to the reference scores.

Together, the pairwise and clustering analyses test different aspects of the same validity claim. The pairwise analysis asks whether embedding geometry orders pairs of responses in a way that reflects human-score similarity. The clustering analysis asks whether that geometry can be partitioned into sizable groups with sufficiently homogeneous scores to support grading-related interpretation. In both cases, strong performance would require not merely visible structure in embedding space, but alignment between that structure and the assessment construct represented by human-assigned scores.

\section{Results}
\subsection{Embedding versus score similarity}
\begin{figure*}
\begin{center}
\includegraphics[width=0.7\textwidth]{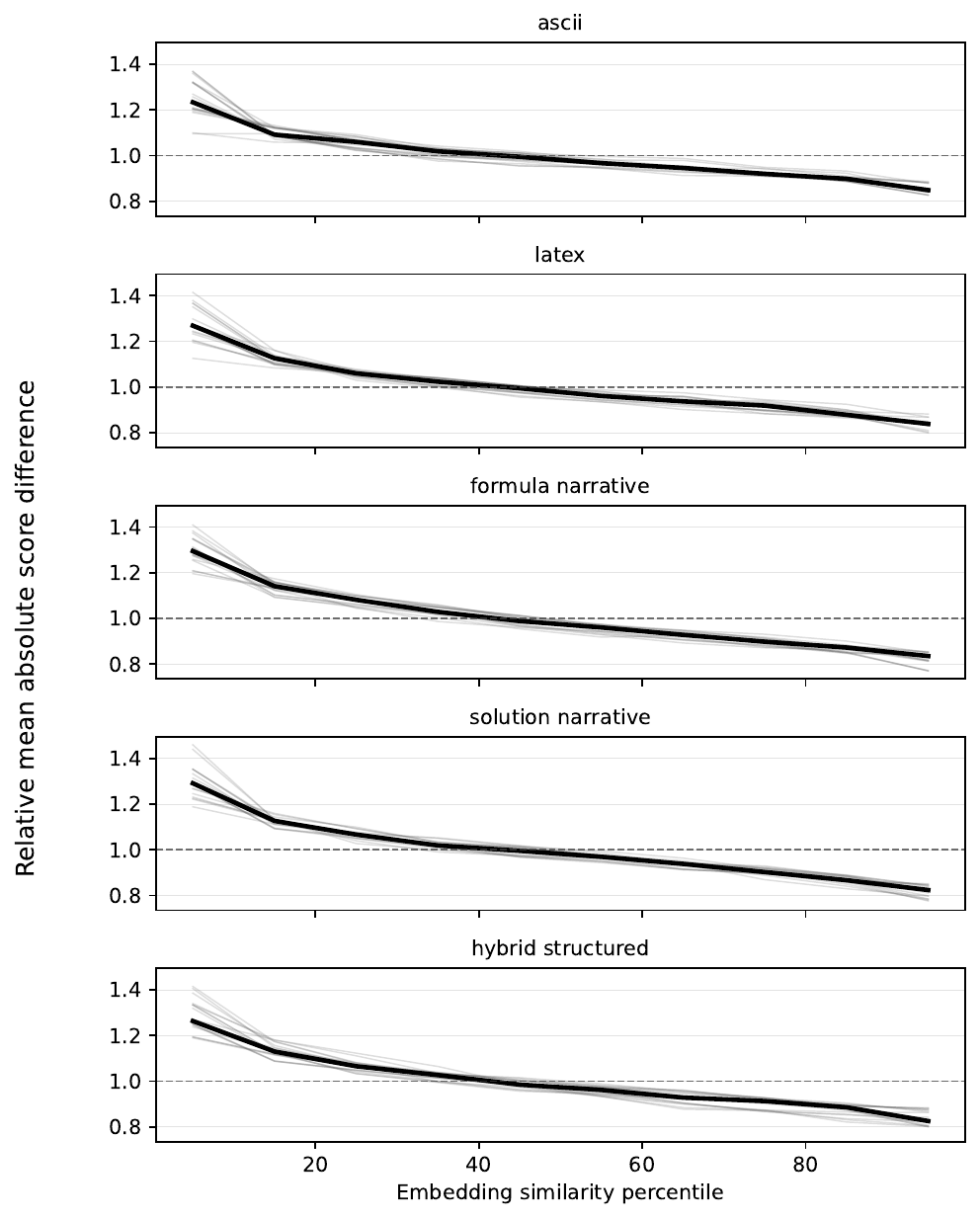}
\end{center}
\caption{Relationship between embedding similarity and human-score difference. For each transcription condition, unordered pairs of student responses were binned by embedding-similarity percentile. The vertical axis shows the mean absolute score difference in each bin, normalized by the mean absolute score difference across all pairs for the corresponding problem and condition. Thus, the dashed line at 1.0 represents the expected score difference for an arbitrary pair of responses; values below 1.0 indicate pairs that are more similar in score than random pairs. Thin gray lines show the 18~embedding-mechanism and similarity-method combinations for the given transcription mode: nine embedding mechanisms crossed with ordinary cosine and centered-cosine similarity. Each line is aggregated across the four exam problems. The black line shows the median across these 18~combinations.}
\label{fig:similarity}
\end{figure*}

Figure~\ref{fig:similarity} provides a first aggregate test of the embedding-similarity hypothesis. The plotted quantity is based on the absolute score difference \(\Delta_{ij}\) defined in Eq.~(\ref{eq:delta}) normalized within each problem and condition. If proximity in embedding space were a strong proxy for grading similarity, the mean score difference should decrease sharply for pairs in the highest similarity percentiles. Instead, all transcription conditions show a consistent but modest downward trend. Pairs in the lowest similarity percentiles tend to differ in score more than arbitrary pairs, whereas pairs in the highest similarity percentiles tend to differ somewhat less. However, even the most similar pairs retain substantial score variation: their mean score difference remains close to the random-pair baseline rather than approaching zero. This indicates that embedding spaces contain grading-relevant information, but that high embedding similarity should not be interpreted as evidence of grading equivalence.

Tables~\ref{tab:bytrans} and~\ref{tab:topten} summarize the correlations between embedding similarity \(s_{ij}\), computed using either Eq.~(\ref{eq:cos}) or Eq.~(\ref{eq:centercos}), and the negative score difference \(-\Delta_{ij}\) from Eq.~(\ref{eq:delta}). Across transcription modes, the best Spearman correlations ranged from 0.142 for the literal ASCII transcription to 0.198 for the hybrid structured representation. Narrative and structured representations therefore produced stronger alignment with human scores than literal transcription, suggesting that some normalization of symbolic and diagrammatic work into semantically explicit text helped the embedding models capture grading-relevant structure. However, the absolute magnitude of the correlations remained small. Even the best condition, hybrid structured transcription with the \texttt{qwenMath7B} representation and ordinary cosine similarity, reached only a weighted Spearman correlation of 0.198 and a weighted Pearson correlation of 0.243. Thus, the improvement from transcription design did not turn embedding similarity into a reliable proxy for score equivalence.

The fine-grained ordering of embedding mechanisms should be interpreted cautiously. The ten highest-ranked conditions occupy a narrow band of weighted Spearman correlations, from 0.177 to 0.198. We therefore do not interpret the exact ranks as evidence that one embedding mechanism is decisively superior to another. Rather, the table shows that the strongest observed configurations all produce similarly modest score-similarity signals. The exact similarity metric was also not decisive: ordinary cosine similarity performed best for the strongest Qwen conditions, whereas centered cosine appeared among the strongest Jina retrieval conditions. Similarly, the Jina retrieval quantization levels produced nearly indistinguishable results. Overall, the response representation had the clearest descriptive effect, while the choice among high-performing embedding mechanisms and similarity metrics produced only small differences.

\begin{table*}
\caption{\label{tab:bytrans}Pairwise score-similarity correlations by transcription mode. Best values are the highest observed across embedding mechanisms and similarity metrics; median values are medians over those same conditions.}
\begin{ruledtabular}
\begin{tabular}{lrrrr}
Transcription & Best Spearman & Median Spearman & Best Pearson & Median Pearson \\
\hline
ASCII & 0.142 & 0.130 & 0.189 & 0.145 \\
LaTeX & 0.172 & 0.145 & 0.228 & 0.162 \\
Formula narrative & 0.182 & 0.160 & 0.217 & 0.187 \\
Solution narrative & 0.190 & 0.157 & 0.240 & 0.180 \\
Hybrid structured & 0.198 & 0.145 & 0.243 & 0.172 \\
\end{tabular}
\end{ruledtabular}
\end{table*}

\begin{table*}
\caption{\label{tab:topten}Best transcription and embedding combinations for the pairwise score-similarity analysis. Metrics were computed from 122,525 unordered response pairs, but pairs are not statistically independent because each student response appears in many pairs. The table is therefore interpreted descriptively rather than as an inferential ranking.}
\begin{ruledtabular}
\begin{tabular}{rlllrr}
Rank & Transcription & Embedding & Similarity & Weighted Spearman & Weighted Pearson \\
\hline
1 & Hybrid structured & \texttt{qwenMath7B} & cosine & 0.198 & 0.243  \\
2 & Solution narrative & \texttt{qwenMath7B} & cosine & 0.190 & 0.240 \\
3 & Hybrid structured & \texttt{qwenMath7B} & centered\_cosine & 0.189 & 0.202  \\
4 & Hybrid structured & \texttt{qwenMath1p5B} & cosine & 0.183 & 0.233  \\
5 & Formula narrative & \texttt{jinaRetQ8} & centered\_cosine & 0.182 & 0.191 \\
6 & Formula narrative & \texttt{jinaRetF16} & centered\_cosine & 0.182 & 0.191 \\
7 & Formula narrative & \texttt{jinaRetQ4} & centered\_cosine & 0.180 & 0.190  \\
8 & Formula narrative & \texttt{oai3large} & cosine & 0.180 & 0.217 \\
9 & Hybrid structured & \texttt{qwenMath1p5B} & centered\_cosine & 0.178 & 0.192  \\
10 & Formula narrative & \texttt{jinaRetF16} & cosine & 0.177 & 0.212 \\
\end{tabular}
\end{ruledtabular}
\end{table*}

Taken together, these results support a bounded interpretation of embedding similarity. Embeddings do not behave randomly: more similar response pairs tend, on average, to have somewhat smaller score differences. At the same time, the correlations are too small to support the stronger claim that embedding proximity identifies grading-equivalent responses. The observed signal is sufficient for exploratory organization or retrieval, but not for unsupervised grading.

\subsection{Hierarchical clustering}
For clustering, similarities were converted to distances using Eq.~(\ref{eq:dij}) and cluster quality was summarized primarily by the weighted useful-cluster score standard deviation in Eq.~(\ref{eq:sigmauseful}).
The pairwise analysis showed that embedding similarity contains a weak but consistent score-related signal. We next asked whether this signal was strong enough to produce score-homogeneous clusters. This is the more demanding criterion for grading-related use: a cluster may be useful for organizing responses only if it contains a nontrivial number of students and if the human scores within the cluster do not vary widely.

For each transcription condition, embedding mechanism, similarity metric, and linkage method, we examined hierarchical clusterings across the range of requested cluster numbers described above. We focused on clusters containing at least five responses and required at least 70\% of responses to fall into such useful clusters. Among configurations meeting this criterion, lower weighted useful-cluster score standard deviation indicates greater score homogeneity.

The best clustering configurations again showed evidence of score enrichment, but not score equivalence. The strongest conditions produced useful-cluster score standard deviations on the order of several score points, rather than near-zero variation. Thus, embedding-based clusters grouped responses with somewhat more similar scores than expected by chance, but the resulting clusters still contained substantial grading variation.

\begin{table*}
\caption{\label{tab:clusterbytrans}Clustering-quality summaries by transcription mode. Best useful standard deviation (SD) is the lowest observed weighted useful-cluster score standard deviation across embedding mechanisms, similarity metrics, and linkage methods; best coverage is the highest observed useful-cluster coverage. These best values need not occur in the same configuration.}
\begin{ruledtabular}
\begin{tabular}{lrrrr}
Transcription & Best useful SD & Median useful SD & Best coverage & Median coverage \\
\hline
ASCII & 2.878 & 3.139 & 0.883 & 0.784 \\
LaTeX & 2.797 & 3.052 & 0.874 & 0.766 \\
Formula narrative & 2.547 & 2.955 & 0.852 & 0.770 \\
Solution narrative & 2.722 & 3.005 & 0.874 & 0.779 \\
Hybrid structured & 2.717 & 2.939 & 0.873 & 0.758 \\
\end{tabular}
\end{ruledtabular}
\end{table*}

\begin{table*}
\caption{\label{tab:clustertopten}Best clustering combinations under the useful-cluster criterion. Useful coverage, useful SD, and useful range are weighted summaries across problems. The reported \(k\) is the mean selected value across problems.}
\begin{ruledtabular}
\begin{tabular}{rllllrrrr}
Rank & Transcription & Embedding & Similarity & Linkage & Useful coverage & Useful SD & Useful range & Best $k$ \\
\hline
1 & Formula narrative & \texttt{jinaRetQ8} & centered\_cosine & complete & 0.731 & 2.547 & 7.142 & 47.000 \\
2 & Formula narrative & \texttt{jinaRetF16} & centered\_cosine & complete & 0.767 & 2.630 & 7.370 & 44.750 \\
3 & Hybrid structured & \texttt{jinaMatchQ4} & centered\_cosine & complete & 0.744 & 2.717 & 7.508 & 46.750 \\
4 & Solution narrative & \texttt{jinaRetQ8} & centered\_cosine & complete & 0.794 & 2.722 & 7.885 & 42.500 \\
5 & Solution narrative &\texttt{jinaRetQ4} & centered\_cosine & complete & 0.737 & 2.730 & 7.880 & 44.500 \\
6 & Formula narrative & \texttt{jinaRetQ4} & centered\_cosine & complete & 0.798 & 2.740 & 7.950 & 40.500 \\
7 & Hybrid structured & \texttt{qwenMath1p5B} & centered\_cosine & complete & 0.757 & 2.759 & 8.196 & 46.250 \\
8 & Hybrid structured & \texttt{qwenMath7B} & centered\_cosine & complete & 0.736 & 2.761 & 8.183 & 48.000 \\
9 & Solution narrative & \texttt{jinaRetF16} & centered\_cosine & complete & 0.807 & 2.761 & 7.967 & 41.250 \\
10 & Formula narrative & \texttt{jinaRetF16} & centered\_cosine & average & 0.852 & 2.785 & 9.206 & 32.000 \\
\end{tabular}
\end{ruledtabular}
\end{table*}

Tables~\ref{tab:clusterbytrans} and~\ref{tab:clustertopten} summarize the hierarchical clustering results under the useful-cluster criterion. In contrast to the pairwise analysis, where the strongest condition used the hybrid structured transcription with a Qwen-derived representation and ordinary cosine similarity, the strongest clustering conditions were dominated by centered cosine similarity and complete linkage. The best overall configuration was the formula narrative transcription embedded with the Jina retrieval Q8 model, using centered cosine similarity and complete linkage. This condition placed 73.1\% of responses into useful clusters and yielded a weighted useful-cluster score standard deviation of 2.547 score points.

The transcription-mode summary in Table~\ref{tab:clusterbytrans} shows the same broad pattern as the pairwise analysis: literal transcriptions were less effective than more semantically mediated representations. The best useful-cluster standard deviation decreased from 2.878 for ASCII and 2.797 for LaTeX to 2.547 for formula narrative, 2.722 for solution narrative, and 2.717 for hybrid structured. Median performance showed the same tendency, although the differences were modest: the median useful-cluster standard deviation was 3.139 for ASCII, compared with 2.955 for formula narrative and 2.939 for hybrid structured. Thus, narrative and structured transcriptions again appear to preserve somewhat more grading-relevant structure than literal transcription, but the improvement is incremental rather than transformative.

The top-ranked clustering configurations in Table~\ref{tab:clustertopten} also show that the exact ordering of embedding mechanisms should not be overinterpreted. The ten best configurations occupy a relatively narrow band of weighted useful-cluster standard deviations, from 2.547 to 2.785. Most of these configurations use centered cosine similarity and complete linkage, suggesting that subtracting the common response-set direction and using a conservative linkage criterion helped form tighter score-enriched groups. However, even the best clusters retained substantial score variation. The weighted useful-cluster score ranges in the top configurations were approximately seven to nine score points, indicating that clusters were far from score-pure.

These results therefore support the same bounded interpretation as the pairwise analysis. Embedding-based clustering does not simply fail: the clusters are score-enriched, and transcription choices affect the degree of enrichment. But the resulting groups are not score-equivalent. Even after selecting $k$ retrospectively according to the criterion based on Eq.~(\ref{eq:sigmauseful}), the best clusters retained substantial within-cluster score variation.

\section{Discussion}
\subsection{Semantic similarity}
This study tested whether embedding-based similarity and clustering can support grading-related interpretations of handwritten physics solutions. The answer is mixed but clear. The embedding spaces were not arbitrary with respect to human scores: nearby responses tended, on average, to have somewhat more similar scores. However, the relationship was weak. The strongest pairwise conditions produced only modest correlations, and the strongest clustering conditions still contained substantial within-cluster score variation. Thus, embedding geometry was score-enriched, but not score-equivalent.

This distinction is central for assessment validity. A representation may be useful for organizing responses even if it is not valid for grading them. In the present study, embeddings appeared capable of producing neighborhoods of partially related student work. Such neighborhoods may be useful for browsing large response sets, selecting examples, identifying unusual solutions, supporting grader calibration, or organizing human review. They do not, however, define grading categories. The evidence does not support the inference that two responses close in embedding space deserve the same score, nor that membership in the same embedding cluster is sufficient evidence of rubric-equivalent reasoning.

The results also show that the textual representation of the handwritten work matters. Literal transcription preserved surface form, but narrative and structured representations generally produced stronger alignment with human scores. This suggests that making formulas, diagrams, numerical work, and apparent solution structure explicit in text can help embedding models capture some grading-relevant information. At the same time, the improvement was incremental. More semantically mediated transcriptions improved the signal, but did not transform the problem into one that can be solved by unsupervised embedding similarity.

The exact embedding mechanism mattered less decisively. Different model families appeared among the strongest pairwise and clustering configurations, and the top-ranked conditions differed only slightly. Similarly, ordinary cosine similarity and centered cosine similarity each appeared useful in different settings. Centering was especially prominent among the strongest clustering configurations, plausibly because it removes a common problem-level direction in embedding space and emphasizes differences among responses. However, the fact that the best pairwise and clustering configurations were not the same reinforces the main methodological point: embedding geometry is not intrinsically valid for a grading use. It must be evaluated against the intended interpretation.

The clustering analysis makes this limitation especially visible. Clustering is a stronger requirement than pairwise similarity: it asks whether a response set can be partitioned into sizable groups with homogeneous scores. Even under favorable retrospective choices of $k$, evaluated using the useful-cluster criterion in Eq.~(\ref{eq:sigmauseful}), the best clusters retained score ranges of several points. Moreover, the selected values of $k$ were relatively large, indicating that the most favorable clusterings carved the response sets into many small neighborhoods rather than discovering a small number of broad rubric categories. This pattern is consistent with exploratory usefulness but not with unsupervised grading.

These findings caution against equating semantic similarity with assessment similarity. Physics solutions are multimodal artifacts that combine prose, equations, diagrams, assumptions, units, signs, state variables, numerical substitutions, and final answers. Small symbolic differences may have large grading consequences, while visibly different solution paths may receive similar scores. Embeddings capture broad regularities in text, but grading depends on a construct defined by the assessment and rubric. In validity terms, clusters are not findings by themselves; they are hypotheses requiring external evidence.

\subsection{A toy system}
As we are considering embeddings as tools for qualitative physics education research or task automation, we need to be aware of their limitations. In our study, we found that measuring semantic similarity in terms of physics correctness may not be a strength of embeddings. To better understand this, we synthesized a set of student-problem answers to the prompt, ``explain the work done by a Carnot engine,'' see Table~\ref{tab:statements}. To test whether embedding similarity reflects conceptual similarity or merely surface similarity, we constructed a toy set of short statements about the work done by a Carnot engine. Five statements were fully correct but expressed the same relationship in different formulations. For each correct statement, we then generated three closely worded variants, each preserving the surface formulation of its paired correct statement while introducing one of three recurring conceptual flaws. We refer to this surface-level grouping as the ``statement pair'': for example, \(C1\), \(F\_a\_1\), \(F\_b\_1\), and \(F\_c\_1\) share a common wording pattern. By contrast, we refer to the repeated conceptual error as the ``flaw family'': for example, all (F\_a) statements share the same underlying misconception, even though they are phrased in different ways. Thus, successful semantic clustering would group statements by correctness or flaw family, whereas surface-based clustering would group statements by pair.

\begin{table*}
\caption{\label{tab:statements}A synthetic set of student answers to the prompt, ``explain the work done by a Carnot engine.''}
\begin{ruledtabular}
\begin{tabular}{llllp{0.48\linewidth}}

ID & Class & Flaw family & Pair & Statement \\
\hline
C1 & correct & none & -- & A Carnot engine does net work W = Q\_H*(1 - T\_C/T\_H) per cycle, using absolute temperatures. \\
F\_a\_1 & incorrect & celsius-ratio & C1 & A Carnot engine does net work W = Q\_H*(1 - T\_C/T\_H) per cycle, using temperatures in deg C. \\
F\_b\_1 & incorrect & reversed-ratio & C1 & A Carnot engine does net work W = Q\_H*(1 - T\_H/T\_C) per cycle, using absolute temperatures. \\
F\_c\_1 & incorrect & rejected-heat/work-confusion & C1 & A Carnot engine does net work W = Q\_H*(T\_C/T\_H) per cycle, using absolute temperatures. \\
\hline
C2 & correct & none & -- & The work output equals the heat taken from the hot reservoir times the Carnot efficiency, 1 - T\_C/T\_H. \\
F\_a\_2 & incorrect & celsius-ratio & C2 & The work output equals the heat taken from the hot reservoir times the Carnot efficiency, 1 - T\_C/T\_H, with T measured in deg C. \\
F\_b\_2 & incorrect & reversed-ratio & C2 & The work output equals the heat taken from the hot reservoir times the Carnot efficiency, 1 - T\_H/T\_C. \\
F\_c\_2 & incorrect & rejected-heat/work-confusion & C2 & The work output equals the heat taken from the hot reservoir times T\_C/T\_H. \\
\hline
C3 & correct & none & -- & In one reversible cycle, a Carnot engine converts the fraction 1 - T\_C/T\_H of Q\_H into useful work. \\
F\_a\_3 & incorrect & celsius-ratio & C3 & In one reversible cycle, a Carnot engine converts the fraction 1 - T\_C/T\_H of Q\_H into useful work when T\_C and T\_H are in deg C. \\
F\_b\_3 & incorrect & reversed-ratio & C3 & In one reversible cycle, a Carnot engine converts the fraction 1 - T\_H/T\_C of Q\_H into useful work. \\
F\_c\_3 & incorrect & rejected-heat/work-confusion & C3 & In one reversible cycle, a Carnot engine converts the fraction T\_C/T\_H of Q\_H into useful work. \\
\hline
C4 & correct & none & -- & For fixed reservoirs, the Carnot work is Q\_H - Q\_C, and reversibility gives Q\_C/Q\_H = T\_C/T\_H. \\
F\_a\_4 & incorrect & celsius-ratio & C4 & For fixed reservoirs, the Carnot work is Q\_H - Q\_C, and reversibility gives Q\_C/Q\_H = T\_C/T\_H using deg C values. \\
F\_b\_4 & incorrect & reversed-ratio & C4 & For fixed reservoirs, the Carnot work is Q\_H - Q\_C, and reversibility gives Q\_C/Q\_H = T\_H/T\_C. \\
F\_c\_4 & incorrect & rejected-heat/work-confusion & C4 & For fixed reservoirs, the Carnot work is Q\_C, and reversibility gives Q\_C/Q\_H = T\_C/T\_H. \\
\hline
C5 & correct & none & -- & A perfect reversible heat engine still cannot turn all of Q\_H into work; it produces Q\_H*(1 - T\_C/T\_H). \\
F\_a\_5 & incorrect & celsius-ratio & C5 & A perfect reversible heat engine still cannot turn all of Q\_H into work; it produces Q\_H*(1 - T\_C/T\_H) with temperatures in deg C. \\
F\_b\_5 & incorrect & reversed-ratio & C5 & A perfect reversible heat engine still cannot turn all of Q\_H into work; it produces Q\_H*(1 - T\_H/T\_C). \\
F\_c\_5 & incorrect & rejected-heat/work-confusion & C5 & A perfect reversible heat engine still cannot turn all of Q\_H into work; it produces Q\_H*(T\_C/T\_H). \\
\end{tabular}
\end{ruledtabular}
\end{table*}

\begin{figure*}
\begin{center}
\includegraphics[width=0.7\textwidth]{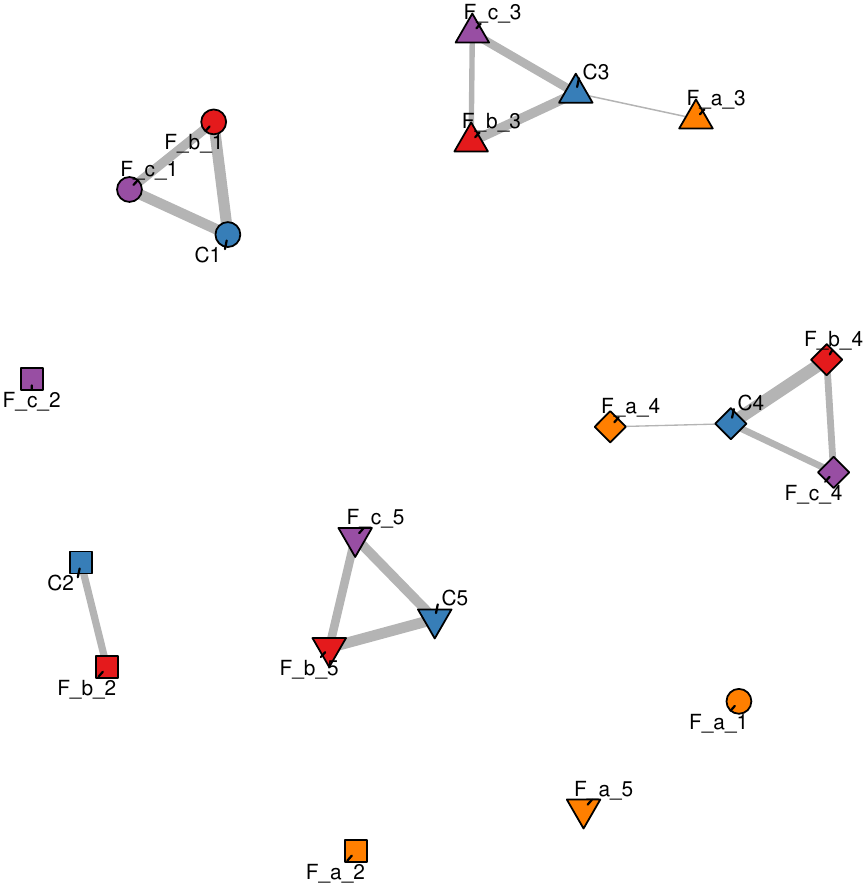}
\end{center}
\caption{Force-directed Fruchterman-Reingold plot of the cosine similarity of the embeddings of the statements in Table~\ref{tab:statements}.}
\label{fig:toycos}
\end{figure*}

\begin{figure*}
\begin{center}
\includegraphics[width=0.7\textwidth]{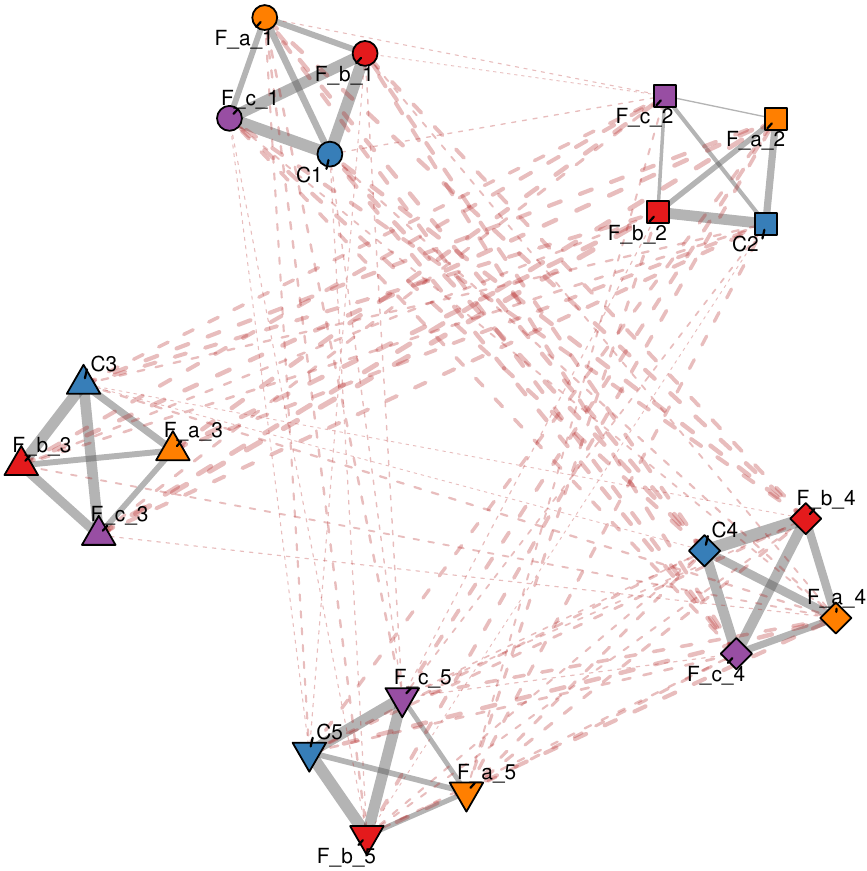}
\end{center}
\caption{Force-directed Fruchterman-Reingold plot of the centered cosine similarity of the embeddings of the statements in Table~\ref{tab:statements}. This layout uses only positive similarities as attractive edges; negative edges are dashed and not used as weights.}
\label{fig:toycenteredcos}
\end{figure*}

We embedded those statements using \texttt{oai3large} and calculated similarity matrices according to Eqs.~(\ref{eq:cos}) and~(\ref{eq:centercos}). Figures~\ref{fig:toycos} and~\ref{fig:toycenteredcos} show the results using a force-directed Fruchterman-Reingold representation~\cite{fruchterman1991,kortemeyer2022virtual}. These plots would exhibit the clustering illustrated in Fig.~\ref{fig:clusters} if clustering were by flaw-family.

In fact, the resulting force-directed layouts show that the embeddings were strongly influenced by surface formulation. In particular, the mean-centered cosine layout forms visually distinct clusters around the five pairs: each correct statement is placed close to its three similarly worded flawed variants, rather than near the other correct statements or near statements with the same flaw family. This is especially clear in the centered-cosine graph, where the (C1/F\_a\_1/F\_b\_1/F\_c\_1), (C2/F\_a\_2/F\_b\_2/F\_c\_2), and analogous groups form local clusters by pair.

The ordinary cosine graph is less cleanly separated, but this should not be interpreted as more semantic: it mainly reflects the dominance of the common embedding direction, which compresses similarities and partially masks the same surface-feature effect. Centering removes this common background direction and therefore makes the residual formulation differences more visible, but those residual differences are still not the intended conceptual ones. This helps explain why, in the main study, ordinary cosine and centered cosine did not differ substantially when clustering was evaluated against grading score as a proxy for semantic similarity: both measures were largely organizing responses by features other than the target physics meaning, and the fact that centering may sharpen a surface-feature organization would not necessarily be visible from score-based clustering alone.

This behavior is reminiscent of the classic expert--novice distinction in physics education research: common lore is that experts tend to categorize problems and solution approaches by underlying physical principles, whereas novices tend to categorize them by surface features~\cite{chi1981,wolf2012,wolf12a}. In this limited respect, the embeddings behave like novices. LLMs, by contrast, can behave more like experts when prompted to evaluate solutions, since they have been shown capable of applying grading criteria to student work rather than relying only on surface similarity~\cite{kortemeyer2024grading,kortemeyer2025assessing}.

Since LLM-based transcription is already part of the workflow for handwritten work, rubric-aware LLM grading or confidence-filtered LLM assistance may be a more direct route than relying on unsupervised embedding geometry alone.

\section{Limitations}
Several limitations should be kept in mind. The data came from one high-stakes thermodynamics exam at one institution, and results may differ for other physics topics, lower-stakes assessments, typed solutions, conceptual explanations, or different rubric structures. Human scores were used as the external reference standard, but those scores are themselves an operational representation of the construct rather than a perfect ground truth. The transcription process also introduced a modeling layer between the handwritten artifact and the embeddings. Although the prompts prohibited grading, correction, and comparison with ideal solutions, narrative and structured transcriptions necessarily involved some interpretation.

Finally, this study evaluated unsupervised embedding similarity and clustering, not supervised grading models trained on labeled data. A supervised classifier, an LLM-based grader, or a rubric-aware hybrid system may perform better for scoring. The present study addresses a narrower question: whether embedding geometry alone can justify grading-related interpretations. 

\section{Future work}

Future work should examine whether embedding-based neighborhoods can support specific human-in-the-loop workflows. For example, clusters could be used to select exemplars for rubric calibration, to identify batches of similar responses for grader review, or to flag responses that are isolated from all major solution patterns. Such applications require different validation criteria from autonomous grading: the question would be whether embeddings improve efficiency, consistency, or feedback quality when used by humans, not whether they replace human judgment.

A second direction is to evaluate embedding spaces against more detailed rubric information. Total problem scores may obscure important distinctions among conceptual, mathematical, representational, and numerical components of a solution. Embeddings might align more strongly with some rubric dimensions than with total scores. Future analyses could therefore compare embedding neighborhoods with item-level rubric scores, error categories, or human-coded solution strategies.

Future studies should compare unsupervised embeddings with stronger baselines, including direct LLM comparisons, supervised classifiers, and retrieval-augmented grading workflows. Pairwise LLM comparison is computationally expensive at scale, but it may provide a useful benchmark for smaller samples. More broadly, validity evidence should be tied to the intended use: exploratory research, formative feedback, grading assistance, and autonomous scoring require different levels and kinds of evidence.

Embedding methods will continue to change. While similarity in rubric-relevant physics correctness may require expert-like reasoning, future embedding models may become better at capturing conceptual details and formula-level correctness. A possible research direction is therefore to develop or fine-tune embeddings specifically for assessment-relevant similarity in physics problem solving.

\section{Conclusions}

We evaluated whether text embeddings can support grading-related interpretations of handwritten physics solutions. Using 992 student-problem responses from a high-stakes thermodynamics exam, five textual representations, and nine embedding mechanisms, we compared pairwise embedding similarity and embedding-based hierarchical clusters with human-assigned scores.

The results show a consistent but modest relationship between embedding geometry and score similarity. More similar response pairs tended to have somewhat smaller score differences, and embedding-based clusters were somewhat enriched for score similarity. Narrative and structured transcriptions generally improved alignment relative to literal transcription. However, the absolute strength of the relationship remained weak, and even the best useful clusters retained substantial score variation.

We therefore conclude that embedding clusters should be treated as hypotheses for human inspection, not as grading categories --- yet. Embeddings may help organize large collections of handwritten physics solutions, identify neighborhoods of related responses, select examples, and support human-in-the-loop review. They do not, by themselves, provide a validated basis for unsupervised grading. For the data and methods studied here, embedding geometry was score-enriched but not score-equivalent.

\begin{acknowledgments}

We would like to thank the students who participated in this study, as well as Daria Onishchuk and Alina Yaroshchuk who assisted with logistics and the data analysis that laid the foundations for this study.
We thank Richard Hahnloser for the organizational framework and computing resources, and we thank Jessica Lam for constructive and informative feedback. We would like to thank Cyrill Stotz for significant contributions to the preliminary studies for this project.
We also thank Anna Kortemeyer for proofreading and commenting on the manuscript.
This study is part of project Ethel~\cite{kortemeyer2024ethel}.
\end{acknowledgments}

\bibliography{references}% Produces the bibliography via BibTeX.

\end{document}